\documentclass[twocolumn,aps,amsmath,amssymb,floatfix,superscriptaddress,prb,longbibliography]{revtex4-1}
\usepackage{graphicx}
\usepackage{dcolumn}
\usepackage{bm}
\usepackage[normalem]{ulem}
\usepackage{xspace}
\usepackage[english]{babel} 
\usepackage{hyperref}
\usepackage{color}
\usepackage[flushleft]{threeparttable}
\def\new{\color{black}}

\def\CGO{CuGa$_2$O$_4$\xspace}

\begin{document}

\title{\new {Disorder-induced spin excitation continuum and spin-glass ground state in the inverse spinel CuGa$_2$O$_4$}}
\author{Zhentao Huang}
\altaffiliation{These authors contributed equally to this work.}
\affiliation{National Laboratory of Solid State Microstructures and Department of Physics, Nanjing University, Nanjing 210093, China}
\author{Zhijun~Xu}
\altaffiliation{These authors contributed equally to this work.}
\affiliation{NIST Center for Neutron Research, National Institute of Standards and Technology, Gaithersburg, Maryland 20899, USA}
\author{Shuaiwei~Li}
\affiliation{Hubei Key Laboratory of Photoelectric Materials and Devices, School of Materials Science and Engineering, Hubei Normal University, Huangshi 435002, China}
\author{Qingchen~Duan}
\affiliation{Tsung-Dao Lee Institute $\&$ School of Physics and Astronomy, Shanghai Jiao Tong University, Shanghai 200240, China}
\author{Junbo~Liao}
\author{Song~Bao}
\author{Yanyan~Shangguan}
\author{Bo~Zhang}
\author{Hao~Xu}
\author{Shufan~Cheng}
\author{Zihang~Song}
\author{Shuai~Dong}
\author{Maofeng~Wu}
\affiliation{National Laboratory of Solid State Microstructures and Department of Physics, Nanjing University, Nanjing 210093, China}

\author{M.~B.~Stone}
\affiliation{Neutron Scattering Division, Oak Ridge National Laboratory, Oak Ridge, TN 37831, USA}

\author{Yiming~Qiu}
\affiliation{NIST Center for Neutron Research, National Institute of Standards and Technology, Gaithersburg, Maryland 20899, USA}

\author{Ruidan~Zhong}
\affiliation{Tsung-Dao Lee Institute $\&$ School of Physics and Astronomy, Shanghai Jiao Tong University, Shanghai 200240, China}

\author{Guangyong~Xu}
\affiliation{NIST Center for Neutron Research, National Institute of Standards and Technology, Gaithersburg, Maryland 20899, USA}

\author{Zhen~Ma}
\email{zma@hbnu.edu.cn}
\affiliation{National Laboratory of Solid State Microstructures and Department of Physics, Nanjing University, Nanjing 210093, China}
\affiliation{Hubei Key Laboratory of Photoelectric Materials and Devices, School of Materials Science and Engineering, Hubei Normal University, Huangshi 435002, China}
\author{G.~D.~Gu}
\email{ggu@bnl.gov}
\author{J.~M.~Tranquada}
\email{jtran@bnl.gov}
\affiliation{Condensed Matter Physics and Materials Science Division, Brookhaven National Laboratory, Upton, New York 11973-5000, USA}
\author{Jinsheng~Wen}
\email{jwen@nju.edu.cn}
\affiliation{National Laboratory of Solid State Microstructures and Department of Physics, Nanjing University, Nanjing 210093, China}
\affiliation{Collaborative Innovation Center of Advanced Microstructures and Jiangsu Physical Science Research Center, Nanjing University, Nanjing 210093, China}


\begin{abstract}
Spinel-structured compounds serve as prototypical examples of highly frustrated systems, and are promising candidates for realizing 
the long-sought quantum spin liquid (QSL) state.
However, structural disorder is inevitable in many real 
QSL candidates
and its impact remains a topic of intense debate. In this work, we conduct comprehensive investigations on CuGa$_2$O$_4$, a spinel compound with significant structural disorder, focusing on its thermodynamic properties and spectroscopic behaviors. No long-range magnetic order is observed down to $\sim$80~mK, as evidenced by magnetic susceptibility, specific heat and elastic neutron scattering measurements. More intriguingly, inelastic neutron scattering experiments reveal a broad gapless continuum of magnetic excitations around the Brillouin zone boundary, {\new resembling the magnetic excitation spectra expected for a QSL. Nevertheless, a spin-freezing transition at $T_{\rm{f}} \approx $ 0.88~K is identified from the cusp in the dc susceptibility curves, where a bifurcation between zero-field-cooling and field-cooling curves occurs. Furthermore, ac susceptibility measurements show a peak close to $T_{\rm{f}}$ at low frequency, which shifts to higher temperature with increasing frequency. These results are evident that CuGa$_2$O$_4$ has spin-glass ground state, consistent with the establishment of short-range order inferred from the specific heat measurements. Collectively, these results illustrate the crucial role of disorder in defining the excitation spectrum out of the disordered ground state.} Our findings shed light onto the broader class of AB$_2$O$_4$ spinels and advance our understanding of the spin dynamics in magnetically disordered systems.
\end{abstract}

\maketitle

\section{Introduction}

Frustrated systems represent a fascinating frontier in condensed matter physics, where competing interactions give rise to quantum fluctuations, resulting in rich emergent phenomena~\cite{wannier_antiferromagnetism_1950,gardner_magnetic_2010,balents_spin_2010}. Of particular interest are quantum spin liquids (QSLs), in which strong quantum fluctuations prevent the establishment of conventional long-range magnetic order, but can also give rise to fractionalized quasiparticle excitations such as spinons~\cite{anderson_resonating_1973,balents_spin_2010,zhou_quantum_2017,wen_experimental_2019,knolle_field_2019,broholm_quantum_2020}. Frustration typically arises from geometrical constraints, as in triangular or kagome lattices~\cite{han_fractionalized_2012,paddison_continuous_2017}, or from competing exchange interactions, as in the honeycomb lattice~\cite{kitaev_anyons_2006,ran_spin-wave_2017,shangguan_one-third_2023}. Within the landscape of frustrated systems, the compounds with a AB$_2$X$_4$ (A = transition metal, B = Al, Ga, Sc; X = O, S) spinel structure stand out as prototypical examples due to their inherent geometrical and magnetic structures~\cite{fiorani_antiferromagnetic_1985,lee_emergent_2002,bergman_order-by-disorder_2007,zaharko_evolution_2010,savary_impurity_2011}. These materials have attracted significant attention for their potential to host novel quantum states~\cite{henley_ordering_1989,bergman_order-by-disorder_2007,tristan_geometric_2005,bernier_quantum_2008}.

The AB$_2$X$_4$ family of compounds crystallizes in the cubic spinel structure (space group Fd$\bar{3}$m), where A-site cations occupy tetrahedral coordination sites, while B-site cations reside at the octahedral coordination sites. If the magnetic ions are located at the A-sites, such compounds are termed  A-site magnetic spinels. Here, the magnetic ions at the A-sites form a diamond-like magnetic lattice, resulting in various exotic phenomena depending on the ratio of the next-nearest-neighbor interaction ($J_{\rm 2}$) to the nearest-neighbor interaction ($J{\rm_ 1}$)~\cite{bergman_order-by-disorder_2007}. This has been the subject of extensive investigations in recent years~\cite{macdougall_kinetically_2011,zaharko_spin_2011,savary_impurity_2011,zaharko_evolution_2010,iakovleva_ground_2015,tristan_geometric_2005}. A notable example is the spiral spin liquid state observed in the highly-frustrated compound MnSc$_2$S$_4$ with $|J_{\rm 2}/J_{\rm 1}|=0.85$~(Refs.~\onlinecite{bergman_order-by-disorder_2007,gao_spiral_2017}). 
CuAl$_2$O$_4$ has also attracted a lot of attention with A-sites nominally occupied by Cu$^{2+}$ with spin-${1}/{2}$~\cite{nirmala_spin_2017,nikolaev_realization_2018}. 
With Cu$^{2+}$ in a tetrahedrally-coordinated site, the 
single 3$d$ hole should be shared among the three-fold degenerate $t_{2g}$ orbitals. 
Taking spin-orbital coupling into account, it might be possible to realize a spin-orbital-liquid ground state~\cite{nirmala_spin_2017}. 
Unfortunately, a study with resonant inelastic x-ray spectroscopy found evidence that the A site has a Jahn-Teller distortion, inconsistent with spin-orbital coupling, and that Cu is also present at the octahedral B site~\cite{huang_resonant_2022}.

Most investigations on the spinel compounds have focused on the normal A-site magnetic spinels~\cite{macdougall_kinetically_2011,zaharko_spin_2011,savary_impurity_2011,zaharko_evolution_2010,iakovleva_ground_2015,tristan_geometric_2005,nirmala_spin_2017,nikolaev_realization_2018,bergman_order-by-disorder_2007,gao_spiral_2017}, while those with inverse spinel structure have received relatively little attention. In reality, site mixing is a common phenomenon in spinel structures, where a fraction of A- and B-site cations interchange positions. This can introduce a high degree of anti-site disorder and, in the extreme case, can lead to the formation of an inverse spinel structure, where all the original A-site magnetic ions sit on the B-sites~\cite{oneill_simple_1983}. The extent of this anti-site disorder can be quantified by the inversion parameter $\delta$, which is expressed as (A$_{1-\delta}$B$_\delta$)[B$_{2-\delta}$A$_\delta$]X$_4$~\cite{fiorani_antiferromagnetic_1985,tristan_geometric_2005,nirmala_spin_2017}. Here, parentheses denote the tetrahedral coordination site, while brackets indicate the octahedral coordination site. {\new The closer the inversion parameter $\delta$ is to 0.5, the stronger is the degree of anti-site disorder.} Note that in the case of CuAl$_2$O$_4$, experiments indicate that $\delta\approx 0.3$~(Refs.~\onlinecite{fregola_cation_2012,cho_dynamic_2020,huang_resonant_2022}). It is noteworthy that such anti-site disorder is typically regarded as an obstacle to the formation of QSLs~\cite{wen_experimental_2019,broholm_quantum_2020}, while others suggest that disorder can, in some cases, play a crucial role in stabilizing the QSL phase~\cite{savary_impurity_2011,furukawa_quantum_2015,savary_disorder-induced_2017}.

In this work, we study \CGO, an inverse spinel compound in which a significant fraction of Cu$^{2+}$ ions occupy octahedral B-sites rather than tetrahedral A-sites. The site disorder has been considered to play a crucial role in determining the physical properties~\cite{sickafus_structure_1999,ndione_control_2014}. To elucidate the underlying physics and clarify the impact of disorder in the magnetic ground state, we perform neutron scattering experiments, complemented by magnetization and specific heat measurements. Our results show that there is no long-range magnetic order down to $\sim$80~mK, but there is a broad gapless continuum of spin excitations as expected for a QSL~\cite{han_fractionalized_2012,shen_evidence_2016,gao_experimental_2019,zeng_spectral_2024}. {\new However, both dc and ac susceptibility measurements reveal a spin-freezing transition at low temperatures, with $T_{\rm f} \approx 0.88$~K determined from the dc susceptibility and a frequency-dependent peak in the real part of the ac susceptibility. This freezing temperature is close to the broad hump around 1~K observed in the specific heat, suggesting the development of short-range magnetic correlations near $T_{\rm f}$. These results evidence that \CGO is a spin glass where disorder plays a crucial role.}

\section{Experimental Details}

{\new Crystals of CuGa$_2$O$_4$ were grown by the floating zone method. To prepare the feed rod, powders of high purity Ga$_2$O$_3$ (99.99\%) and CuO (99.99\%) were mixed in the ratio of 1.085:1 (where the excess Ga allows for evaporation during the single-crystal growth process), ground, and calcined for 48~h at 1150$^\circ$C. The reground powders were placed in a rubber tube and hydrostatically pressed under 6000~kg/cm$^2$. The pressed rods, with 8~mm diameter, were sintered for 72~h at 1350$^\circ$C for use as feed rods. The single-crystal rod  was grown at a velocity of 0.4~mm/h in 1 bar oxygen flow. The grown rod was examined under an optical polarization microscope, where differently oriented crystals appear with different colors. Single crystals were identified and extracted based on these images.}
To prepare powder samples for the x-ray diffraction (XRD) characterization, single crystals were ground into fine powders. The XRD data were collected at room temperature using an x-ray diffractometer (SmartLab SE, Rigaku) with Cu-$K_{\alpha}$ radiation ($\lambda$~=~1.54~\AA).
The diffraction scan covered the angular range from 10$^{\circ}$ to 120$^{\circ}$ with a step size of 0.02$^{\circ}$ and a scanning rate of 10$^{\circ}$/min. The collected XRD data were then subject to Rietveld refinement using the Fullprof software suite to extract structural parameters, including the occupancy of Cu$^{2+}$ and Ga$^{3+}$ ions at the A and B sites and the inversion parameter $\delta$.

Single crystals were aligned using a Laue x-ray diffractometer and then cut using a wire cutting machine. The dc magnetic susceptibility above 2~K was measured in a commercial Quantum Design physical property measurement system (PPMS), while those below 2~K were measured in a commercial Quantum Design magnetic property measurement system (MPMS) equipped with a He-3 refrigerator module. The ultralow-temperature specific heat measurements were performed in a PPMS equipped with a dilution refrigerator. {\new AC magnetic susceptibility measurements were performed using both the ACDR (AC Dilution Refrigerator) and ACMS (AC Measurement System) modules of the Quantum Design PPMS. The same sample was measured in both setups to ensure consistency. The ACDR module was used to access the ultralow-temperature regime down to $\sim$50~mK, while the ACMS measurements were carried out above 1.9~K under identical excitation conditions ($2\times10^{-4}$~T). To ensure the accuracy of the low-temperature data, background correction was applied to the ACDR results using the ACMS data as a reference in the overlapping temperature range (from 1.9~K to 4.0~K).}

Neutron scattering experiments were performed on MACS~\cite{rodriguez_macsnew_2008} with proposal ID 26266 and SPINS at the NIST Center for Neutron Research, and SEQUOIA~\cite{granroth_sequoia_2010} at Oak Ridge National Laboratory with proposal No.~IPTS-24420. On MACS, the incident energy $E_{\rm i}$ was fixed to 5~meV. The measurements were conducted by rotating the samples at 0.3, 15, and 100~K, respectively. In the time-of-flight experiment at SEQUOIA, the measurements were made at the base temperature of 5~K with two $E_{\rm i}$'s of 20 and 40~meV, with a Fermi chopper frequency of 300 Hz, and a scan of rotation about the sample's vertical axis over a range of 154$^\circ$.

Measurements were also performed at $T=100$~K and used for a quantification of the sample background as described later in the text. The wave vector $\bm{Q}$ was expressed in reciprocal lattice units (r.l.u.) as $(H, K, L)$, where $a^{\text{*}} = b^{\text{*}} = c^{\text{*}} = {2\pi}/{a}$ with $a = 8.27$~{\AA} in the space group Fd$\overline{3}$m. The sample used for the neutron scattering measurements consisted of 3 rods with a total mass of 19.8~g, and was aligned in the ($HK0$) scattering plane.

\section{Results}
\subsection{Crystal structure and x-ray diffraction}

\begin{figure}[htb]
	\centering
	\includegraphics[width=\linewidth]{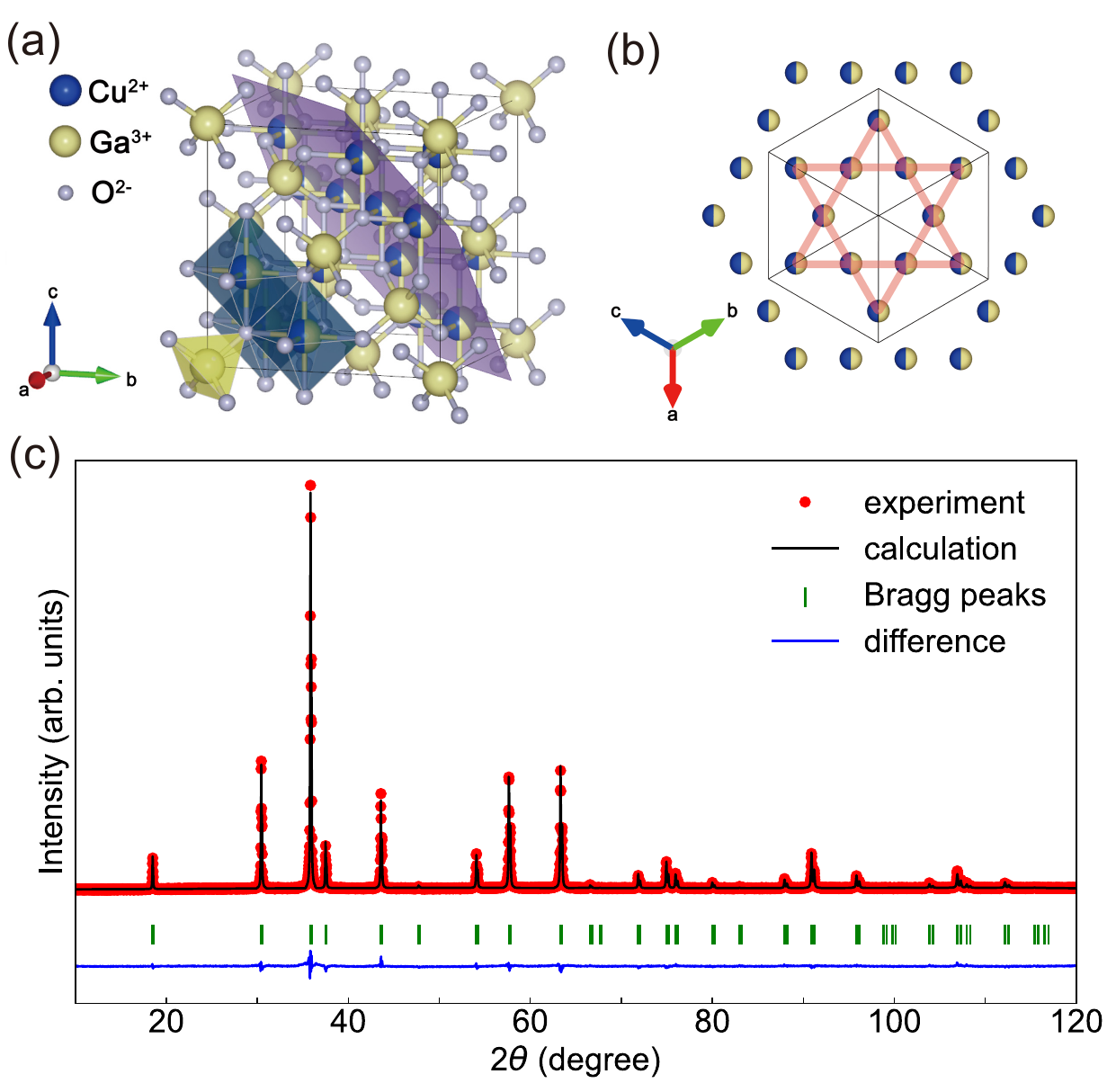}
	\caption{(a) Schematic illustration of the crystal structure of \CGO, drawn using VESTA~\cite{momma_vesta_2011}. The tetrahedral environment at the A site and the octahedral environment at the B site are also illustrated. (b) A typical kagome layer viewed along the [111] direction as illustrated by the purple shaded region in (a). The sphere with mixed blue and yellow colors represents the random occupation of Cu$^{2+}$ and Ga$^{3+}$ ions at these sites. Black lines indicate the boundary of CuGa$_2$O$_4$'s unit cell. (c) Rietveld refinement results for the powder XRD data collected at room temperature. Red circles and black solid line indicate the measured data and calculated results upon Rietveld refinement, respectively. Green ticks denote Bragg peak positions of \CGO with space group F$\overline{3}$dm. The blue solid line represents the difference between the experiment and calculations for this compound.}
	\label{fig1}
\end{figure}

We begin with a detailed description of the structure. Figure~\ref{fig1}(a) illustrates the crystal structure of \CGO, an inverse spinel compound with major Cu$^{2+}$ ions located at the B sites, as indicated by previous powder diffraction studies~\cite{lopitaux_etude_1976,gonzalez_x-ray_1985}. The A and B sites have tetrahedral and octahedral coordinations, respectively. When viewed along the [111] direction, the ions at the B sites form a kagome plane [highlighted by the purple plane  in Fig. 1(a)]. This is a well-known geometrically frustrated network as shown in Fig.~\ref{fig1}(b). Figure~\ref{fig1}(c) presents the Rietveld refinement results for the powder XRD data of \CGO collected at room temperature. The refinement process was performed under the assumption that there is a small copper vacancy due to vapor loss during the growth procedure, no oxygen vacancy, and a slight crystalline preferred orientation in the powder sample. The fitting results show excellent agreement with the experimental data, as indicated by the minimal residual difference line. The goodness of fit, $\chi^2$, is 3.86, with final refinement parameters $R_{\mathrm{p}} \approx 4.71\%$ and $R_{\mathrm{wp}} \approx 6.52\%$. It gives rise to the cubic lattice constant $a = 8.303$~\AA. This parameter is in good agreement with those reported in the existing literature~\cite{petrakovskii_spin-glass_2001}. The detailed refinement results are summarized in Table~\ref{tab:atomic_positions}.  The refined site occupancies of Cu$^{2+}$ ions yield a total net occupancy of 0.98 and an inversion parameter, $\delta \approx 0.76$, 
qualitatively consistent with previous studies~\cite{lopitaux_etude_1976,gonzalez_x-ray_1985}.

The majority of Cu$^{2+}$ ions occupy the B sites, but fill a minority of these sites, so that the magnetic kagome network is quite disordered. The nearest distance between the B-site ions is 2.935~\AA, while the nearest-neighbor and next-nearest-neighbor distances between ions at the A sites are 3.595~\AA~and 5.871~\AA, respectively. The shortest distance between the A and B sites is 3.442~\AA. The relatively short distance  between nearest-neighbor B sites indicates that the magnetic interactions between B-site Cu$^{2+}$ nearest neighbors likely play a dominant role in determining the magnetic properties of this compound.

\begin{table}[hbt]
    \caption{\label{tab:atomic_positions} Refined atomic positions, occupancy (Occ.), and isotropic atomic displacement parameter \( U_{\text{iso}} \). The positions of Cu and Ga ions are fixed based on the symmetry while those of O are refined.}
    \begin{ruledtabular}
    \begin{tabular}{lcccccc}
        Atom & Site & \( x \) & \( y \) & \( z \) & Occ. & \( U_{\text{iso}} \) (\AA$^2$) \\
        \hline
        Cu  (A-site)& \( 8a \)  & 0.00000  & 0.00000  & 0.00000  & {\new 0.24(2)}  & 0.015  \\
        Ga (A-site) & \( 8a \)  & 0.00000  & 0.00000  & 0.00000  & {\new 0.76(8)}  & 0.015  \\
        Ga (B-site)& \( 16d \) & 0.62500  & 0.62500  & 0.62500  & {\new 0.61(3)}  & 0.015  \\
        Cu (B-site) & \( 16d \) & 0.62500  & 0.62500  & 0.62500  & {\new 0.37(3)}  & 0.015  \\
                O  & \( 32e \) & 0.37850  & 0.37850  & 0.37850  & 1.000  & 0.020  \\
            \end{tabular}
    \end{ruledtabular}
\end{table}

\subsection{Magnetic susceptibility and specific heat \label{subsec:ss}}

\begin{figure}[h!]
	\centering
	\includegraphics[width=0.9\linewidth]{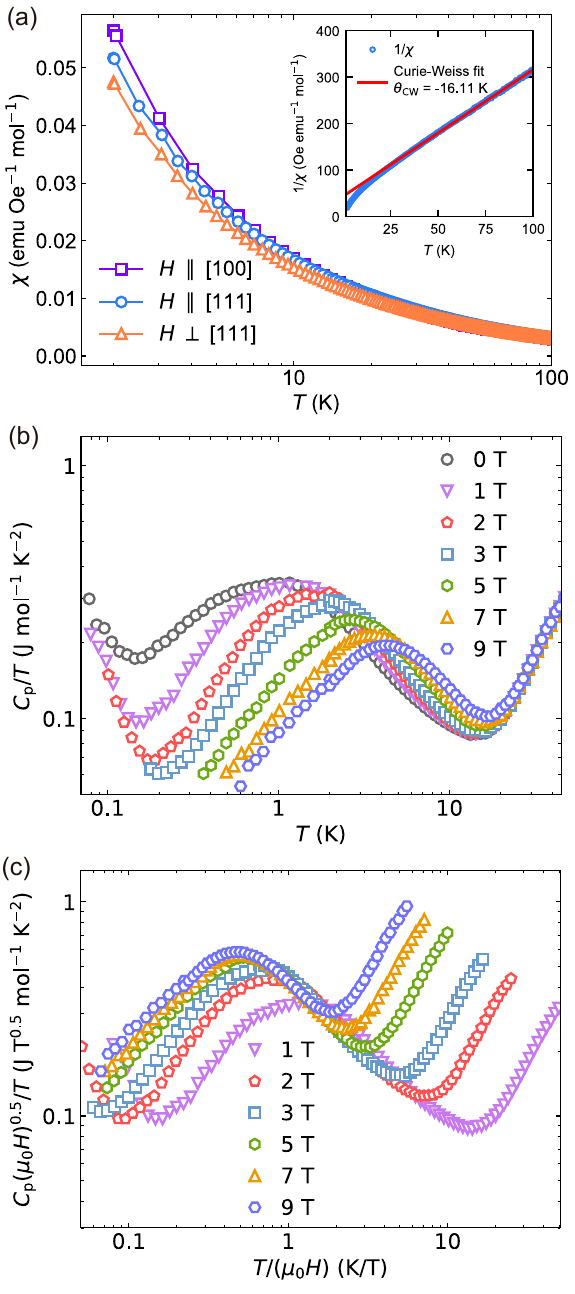}
	\caption{(a) Susceptibility measured with an external magnetic field of 0.1~T applied parallel to the [100], [111] directions, and perpendicular to the [111] direction. No significant difference is observed in the paramagnetic phase. The inset shows the Curie-Weiss fit in the temperature range from {\new 25 to 100~K}, which yields a negative Curie-Weiss temperature of {\new $-16.11$~K}, indicating antiferromagnetic interactions in \CGO. (b) Specific heat measured at different applied magnetic fields. (c) Scaling analysis of the specific heat at different fields ranging from 1 to 9~T.}
	\label{fig2}
\end{figure}

To investigate the magnetic properties of \CGO, magnetic susceptibility and specific heat  have been measured. Figure~\ref{fig2}(a) shows the susceptibility curves measured with an external field of 0.1~T applied along the [100] and [111] directions, and perpendicular to the [111] direction. All three curves show no sign of phase transitions as the temperature approaches the lowest value of 2~K. It is noted that there is a slight bifurcation observed at low temperatures among the susceptibility curves measured with different field directions, indicating the presence of weak magnetic anisotropy. We fit the data with the field applied along the [111] direction using the Curie-Weiss law: $1/\chi = (T - \theta_{\rm{CW}})/\rm{C}$
over the temperature range from {\new 25 to 100~K}, as shown in the inset of Fig.~\ref{fig2}(a). This yields a negative Curie-Weiss temperature of {\new $\theta_{\rm{CW}} \approx -16.11$~K}, indicating dominant antiferromagnetic interactions between Cu$^{2+}$ ions. Note that this is an order of magnitude lower than the value of $-137$~K reported for CuAl$_2$O$_4$ \cite{nirmala_spin_2017}. The effective moment extracted from the fitted Curie-Weiss constant is approximately {\new 1.71} $\mu_{\rm{B}}$/Cu$^{2+}$, which is comparable to the theoretical value of 1.73 $\mu_{\rm{B}}$/Cu$^{2+}$ expected for spin 1/2.

Specific heat measurements down to $\sim80$~mK at various applied fields along the [100] direction are shown in Fig.~\ref{fig2}(b). The data are plotted as $C_p/T$ {\it vs.}\ $T$ with a logarithmic scale. We do not attempt to subtract the phonon contribution, as it is expected to be negligible at low temperatures~\cite{helton_spin_2007,yamashita_thermodynamic_2008}. Despite the nuclear Schottky contribution for $T\lesssim0.2$~K, a broad hump is observed around 1~K at zero field, which shifts to higher temperatures with increasing applied magnetic fields. Such behavior is typically attributed to the establishment of short-range correlations. This is a common feature in frustrated systems, and likewise often found in QSL candidates~\cite{helton_spin_2007,huang_heat_2021,ma_possible_2024}.

We note that due to the disorder resulting from the inter-occupancy between the A and B sites, a random-singlet state may emerge for a frustrated system~\cite{kimchi_valence_2018}. To assess this possibility, we perform further analysis upon these data. A random-singlet state is proposed to exhibit a scaling behavior, where the specific heat data at different applied fields should collapse into a single curve~\cite{kimchi_scaling_2018,peng_dynamical_2024}. This scaling follows the functional form
\(
\frac{C[H, T]}{T} \sim \frac{1}{H^{\gamma}} \left[\frac{T}{H}\right]
\)
described in Ref.~\onlinecite{kimchi_scaling_2018}. In Fig.~\ref{fig2}(c), we 
present our data in this form with the parameter $\gamma = 0.5$. Ignoring the phonon-caused upturn for $T/H\gtrsim1$, we find the data do not collapses into a single curve with a peak as described by the above scaling function, and we have found no improvement by varying $\gamma$.
This is different from the case of herbertsmithite ZnCu$_3$(OH)$_6$Cl$_2$, where the scaling works well~\cite{han_correlated_2016,kimchi_scaling_2018}. These results indicate that the random singlet is not likely to be the ground state for \CGO.

\subsection{Neutron scattering}

\begin{figure*}[htb]
	\centering
	\includegraphics[width=0.9\linewidth]{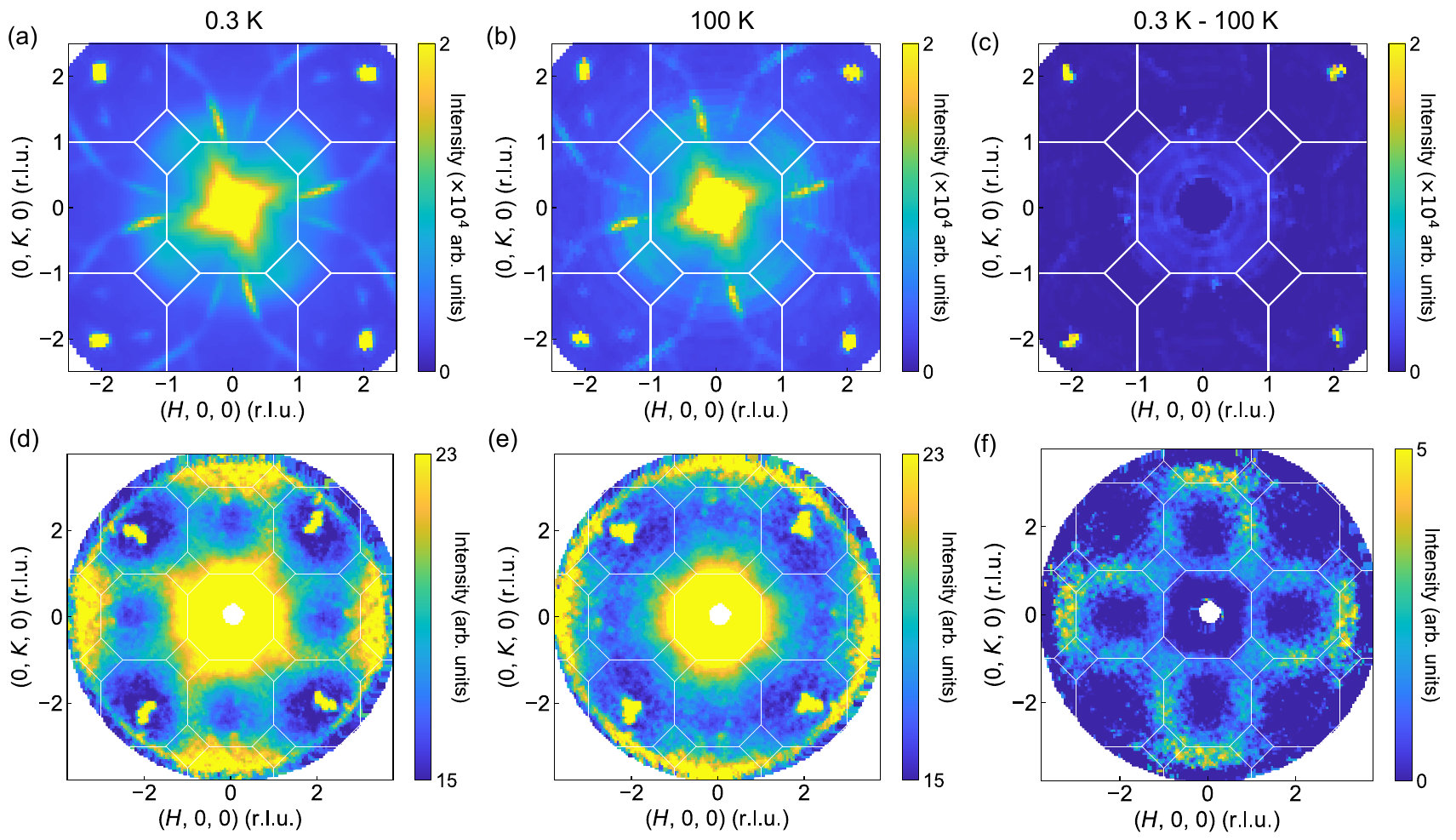}
	\caption{(a), (b) Elastic neutron scattering data of \CGO in the ($H$, $K$, 0) plane, measured at 0.3 and 100~K, respectively. (c) Contour map obtained by subtracting the 100-K data from the 0.3-K data. White lines indicate the Brillouin zone boundaries. (d), (e) Inelastic neutron scattering data with an energy transfer of $\Delta E = 2$~meV in the ($H$, $K$, 0) plane, measured at 0.3 and 100~K, respectively. (f) Contour map obtained by subtracting the 100-K data from the 0.3-K data. To improve data statistics, all data have been folded according to the fourfold rotational symmetry.}
	\label{macs}
\end{figure*}

\begin{figure*}[htb]
	\centering
	\includegraphics[width=0.9\linewidth]{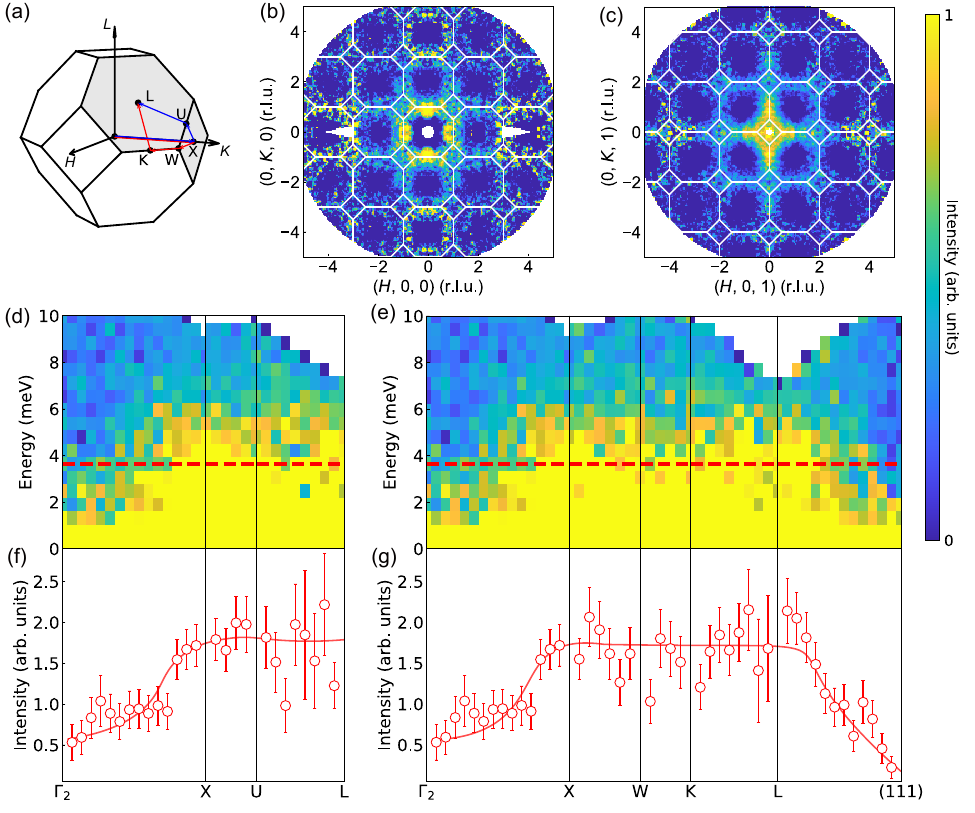}
	\caption{(a) The first Brillouin zone of \CGO, with high-symmetry points labeled. (b), (c) Cuts of time-of-flight neutron scattering data in the ($H$, $K$, 0) and ($H$, $K$, 1) planes, respectively. The energy transfer is 3~meV, with an integrated energy range of $\pm$1~meV and an integrated $L$ range of $\pm$0.25~r.l.u.. To improve data statistics, the original data have been folded along the [110], [-110], and [010] axes. (d), (e) Low-energy spectra along high-symmetry paths denoted in (a) by the blue and red lines, respectively, generated using the \text{spaghetti\_plot} method in \texttt{Horace}\cite{ewings_horace_2016}. Data are integrated over 0.2 \AA$^{-1}$ in perpendicular $\bm{Q}$-directions. (f), (g) Constant-energy cuts at 3.6~meV with energy integrated over 0.6~meV, as indicated by the red dashed lines in panels (d) and (e). Error bars indicate one standard deviation ($\pm\sigma$) of the data. The curves over the data points are guides to the eye.}
	\label{tof}
\end{figure*}

\begin{figure}[htb]
	\centering
	\includegraphics[width=0.95\linewidth]{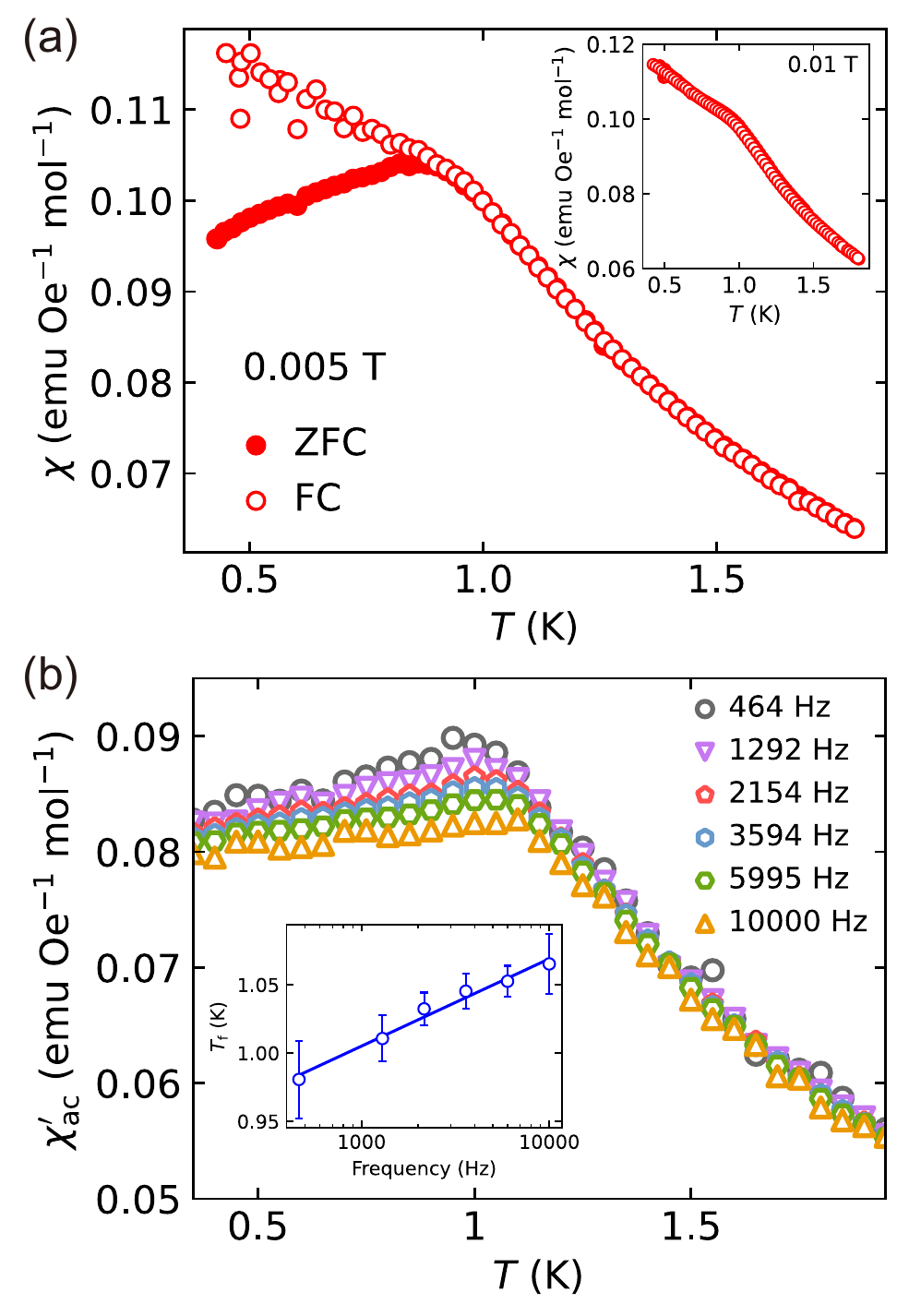}
	\caption{(a) Temperature dependence of the magnetic susceptibility measured under zero-field-cooling (ZFC) and field-cooling (FC) conditions with a small magnetic field of 0.005~T. A bifurcation at $T_{\rm{f}} \approx$ 0.88~K is observed. The inset shows a similar measurement but with a 0.01 T field, where no bifurcation between ZFC and FC curves  observed. {\new (b) Real part of the ac magnetic susceptibility $\chi'_{\mathrm{ac}}$ of CuGa$_2$O$_4$ measured at various frequencies under an excitation field of $2\times10^{-4}$~T. A clear shift of the freezing temperature $T_{\rm{f}}$ toward higher temperatures is observed with increasing frequency, consistent with the spin-glass behavior. The inset shows the frequency dependence of $T_{\rm{f}}$ plotted on a semi-logarithmic scale, where $T_{\rm{f}}$ is defined as the temperature corresponding to the peak in $\chi'_{\mathrm{ac}}(T)$. The solid line is a guide to the eye.}}
	\label{fig5}
\end{figure}

To further elucidate the ground state of \CGO, we conducted both elastic and inelastic neutron scattering (INS) experiments with the spectrometer MACS using single crystals. Figures~\ref{macs}(a) and ~\ref{macs}(b) show elastic neutron scattering intensity maps in the $(H, K, 0)$ plane, measured at 0.3 and 100~K, respectively. Compared to 100~K, no additional peaks appear at 0.3~K. To enhance visibility and identify any potential magnetic peaks, we subtract the 100-K data from the 0.3-K data; the results are shown in Fig.~\ref{macs}(c). The resulting difference reveals no authentic signals but only weak residual intensity, confirming the absence of long-range antiferromagnetic order. These findings suggest that \CGO remains magnetically disordered down to at least 0.3~K.

Here, we also probe the low-energy magnetic excitations of \CGO. Figures~\ref{macs}(d) and ~\ref{macs}(e) display INS intensity maps at 2~meV, measured at 0.3 and 100~K, respectively. At 0.3~K, we observe a broad, ring-like continuum of magnetic excitations concentrated around the Brillouin zone boundary, especially for the second Brillouin zone. In contrast, these excitation signals vanish at 100~K, confirming its magnetic origin. To highlight these excitations, we subtract the 100-K data from the 0.3-K data, yielding the net excitation spectra shown in Fig.~\ref{macs}(f). Similar spectra along the Brillouin zone are observed at other energies. The diffuse, continuum-like nature of these excitations is reminiscent of those observed in other QSL candidate, such as herbertsmithite and YbMgGaO$_4$~\cite{han_fractionalized_2012,han_correlated_2016,shen_evidence_2016}, where spin fluctuations persist in the absence of a long-range order.

To further confirm the authenticity of these observed magnetic excitation signals, we conducted INS experiments using the time-of-flight spectrometer SEQUOIA. Figures~\ref{tof}(b) and \ref{tof}(c) show scattering intensity maps in the $(H, K, 0)$ and $(H, K, 1)$ planes at 3~meV, with an energy integrated width of $\pm 0.1$~meV. It reveals broad excitation modes similar to those observed in Fig.~\ref{macs}, confirming the non-dispersive nature at low energies. By analyzing the dispersion along high-symmetry directions in the Brillouin zone, as indicated in Fig.~\ref{tof}(a), we observe the diffusive low-energy excitations with weak momentum $\bm{Q}$ dependence~[Figs.~\ref{tof}(d) and \ref{tof}(e)], in stark contrast with well-defined spin waves. The excitation spectra remain continuous over a broad energy range, extending up to approximately 6~meV, with no spin gap observed within the instrumental resolution. To provide further insight into the momentum dependence of the excitation spectra, we present constant-energy cuts at 3.6~meV with energy $E$ integrated over 0.6 meV and momentum $\bm{Q}$ integrated over 0.2 \AA$^{-1}$ in perpendicular $\bm{Q}$-directions, which are shown in Figs.~\ref{tof}(f) and \ref{tof}(g). The scattering intensity increases as $\bm{Q}$ moves away from the $\Gamma_2$ point towards the high-symmetry points in the first Brillouin zone. The intensity is broadly distributed across the Brillouin zone, with no sharp, well-defined magnon modes.

\subsection{{\new Sub-Kelvin dc and ac susceptibility}}

The magnetic susceptibility, ultralow-temperature specific heat, and {\new elastic} neutron scattering results all {\new point to the absence of long-range magnetic order} in \CGO. In particular, {\new INS reveals a broad, gapless excitation continuum along the Brillouin zone boundary, reminiscent of dynamic spin fluctuations observed in some QSL candidate materials. However, considering the significant anti-site disorder between Cu$^{2+}$ and Ga$^{3+}$ ions in this inverse spinel structure (see Table~\ref{tab:atomic_positions}), the impact of disorder must be carefully evaluated~\cite{zhou_quantum_2017,wen_experimental_2019}.} As discussed in Sec.~\ref{subsec:ss}, we believe the random-singlet state is less likely according to the scaling analysis of the specific heat data. Now we turn to assess the possibility of a spin glass, another prevalent case in frustrated systems with strong disorder~\cite{binder_spin_1986}. In Fig.~\ref{fig5}, we plot the {\new dc and ac} susceptibility measured below $T = 2$~K. 
The {\new dc susceptibility data in Fig.~\ref{fig5}(a)} show that there is a cusp at $\sim$0.88~K, where the zero-field-cooling (ZFC) and field-cooling (FC) susceptibility curves measured at 0.005~T exhibit a clear bifurcation. This indicates that the system is in fact a spin glass, consistent with previous reports on this compound~\cite{petrakovskii_spin-glass_2001,fenner_zero-point_2009}, albeit with a lower freezing temperature $T_{\mathrm{f}}$ than found in prior measurements. When the applied field is increased to 0.01~T as shown in the inset of {\new Fig.~\ref{fig5}(a)}, the ZFC and FC curves overlap closely, suggesting the destruction of spin clusters in a spin glass.

{\new To confirm the nature of a spin-glass ground state, we also perform frequency-dependent ac magnetic susceptibility measurements. Figure~\ref{fig5}(b) shows the real part of the ac susceptibility $\chi'_{\mathrm{ac}}(T)$ with the driving frequency ranging from 464~Hz to 10000~Hz within 0.5~K and 2~K. There is a clear peak around 1~K and it displays a characteristic frequency dependence, consistent with the spin-glass ground state~\cite{binder_spin_1986,mydosh_spin_2015}. To quantitatively extract the $T_{\mathrm{f}}$ values at various frequencies, we fit the $\chi'_{\mathrm{ac}}(T)$ peaks with the temperature range of 0.6~K to 1.5~K by using an empirical asymmetric Lorentzian function:
\begin{equation}
\chi_{\rm ac}(T) = \chi_0 + \frac{A}{1 + \left( \frac{T - T_f}{w \cdot (1 + \lambda \cdot \mathrm{sgn}(T - T_f))} \right)^2},
\end{equation}
where $\chi_0$, $A$, $w$, and $\lambda$ represent the background, peak amplitude,  intrinsic half-width, and asymmetry of the peak profile, respectively. The sign function $\mathrm{sgn}(T - T_f)$ ensures the asymmetry is applied correctly across the peak center. It gives rises to a value of $T_{\mathrm{f}}$ = 0.98~K for 464~Hz, which is quite close to that of 0.88~K observed in dc magnetic susceptibility measurements. The inset of Fig.~\ref{fig5}(b) shows the extracted $T_{\rm f}$ values as a function of frequency on a semi-logarithmic scale. The relative shift parameter, defined as $\Delta P = \Delta T_{\rm f} / [T_{\rm f} \Delta \log(f)]$, is estimated to be $0.061(9)$, in good agreement with the values reported in other canonical insulating spin-glass systems~\cite{mehlawat_fragile_2015,mydosh_spin_2015}.}

\section{Discussions and Conclusions}

Via comprehensive measurements using XRD, susceptibility, specific heat, and neutron scattering techniques, we have shown that although the frustrated inverse-spinel compound \CGO exhibits spin-liquid-like behaviors, its ground state is indeed a disorder-induced spin glass. According to the scaling analysis, we do not observe a collapse of the specific heat data, and therefore we do not consider the ground state to be the random-singlet state~\cite{kimchi_scaling_2018,bairwa_quantum_2025}. We note that in the A-site CuAl$_2$O$_4$, 
the tetrahedral symmetry of the dominant magnetic sites is distinct from the present case and could lead to a different ground state~\cite{nirmala_spin_2017}. In our case, however, Cu ions mainly occupy the B sites. Under the octahedral crystal-field environment, the $e_g$ orbitals, rather than the $t_{2g}$ orbitals, are active. Such a system is expected to undergo a Jahn-Teller distortion which lifts the degeneracy of the $e_g$ orbitals~\cite{jahn_stability_1937} and prevents the formation of an spin-orbital liquid. Therefore, we do not consider the spin-orbital-liquid ground state here either.

In magnetically frustrated systems, the role of disorder has been extensively studied~\cite{murayama_effect_2020,huang_heat_2021,kao_disorder_2021}. Now, it has been recognized that disorder can often render spin-liquid-like features in a frustrated system~\cite{furukawa_quantum_2015}. A representative example is the QSL candidate herbertsmithite, where the site mixing between magnetic Cu$^{2+}$ and nonmagnetic Zn$^{2+}$ makes the magnetic excitations mimic those expected for a QSL~\cite{norman_colloquium_2016}. Tm$_3$Sb$_3$Zn$_2$O$_{14}$, another material with a kagome structure, which suffers from site-mixing disorder between magnetic Tm$^{3+}$ and nonmagnetic Zn$^{2+}$, also exhibits properties mimicking characteristics of a QSL state~\cite{ding_possible_2018,ma_disorder-induced_2020}. Similarly, a disorder-induced spin excitation continuum has been recently observed in CoZnMo$_2$O$_7$~(Ref.~\onlinecite{gao_disorder-induced_2023}). Besides the examples discussed above, in which the anti-site disorder arises from the direct transposition between magnetic and nonmagnetic ions, there are also cases without  disorder in the magnetic sites, 
such as YbMgGaO$_4$ and YbZnGaO$_4$~(Refs.~\onlinecite{shen_fractionalized_2018,ma_spin-glass_2018, ma_disorder-induced_2021}). In these materials, disorder stems from the random distribution of nonmagnetic Mg$^{2+}$ (or Zn$^{2+}$) ions and Ga$^{3+}$, creating a spatially fluctuating charge environment. Here disorder is believed to influence the ground state by randomizing the pseudo-dipolar interactions of the magnetic ions, thereby promoting QSL-like behaviors~\cite{zhu_disorder-induced_2017}. In these samples, the continuous spin excitation spectra observed in the absence of long-range order closely resemble the behaviors seen in \CGO. These studies highlight that disorder, whether magnetic or nonmagnetic in nature, can induce the spin-liquid-like behaviors in geometrically frustrated materials.

{\new Finally, we would like to compare our results with those reported in Ref.~\onlinecite{petrakovskii_spin-glass_2001}, where the spin-glass behavior in CuGa$_2$O$_4$ was first identified. In that work, the spin-freezing transition was observed at $T_{\mathrm{f}} \approx 2.5$~K based on magnetization, specific heat, and $\mu$SR measurements. Our sample shows a lower freezing temperature of $T_{\mathrm{f}} \approx 0.88$~K, which rapidly disappears with the field increasing to 0.01~T. We note that the Curie-Weiss temperature $\theta_{\rm CW}$ also exhibits some discrepancy between these two works (-8~K in Ref.~\onlinecite{petrakovskii_spin-glass_2001} {\it vs.} -16.11~K in this work). This discrepancy may arise from the sensitivity of $\theta_{\rm CW}$ to the temperature range used in the Curie-Weiss fitting. For instance, a significantly larger value of $\theta_{\rm CW} \approx -40$~K has been reported when the fitting is performed over a higher temperature range above 130~K~\cite{bullard_magnetic_2021}. Moreover, our ac susceptibility measurements show a larger frequency shift parameter ($\Delta P \approx 0.061(9)$) compared to $\sim$0.026 in Ref.~\onlinecite{petrakovskii_spin-glass_2001}. We attribute these differences to the different degrees of disorder in these samples, since the single crystals were prepared by distinct methods. They obtained the samples via flux method while we grew single crystals using the floating zone method with a higher temperature. Considering the comparable ionic radii for Cu$^{2+}$ and Ga$^{3+}$ in both tetrahedral and octahedral coordinations~\cite{shannon_revised_1976}, subtle variation in heat-treated temperature by different means could affect the degree of Cu/Ga site mixing and then the extracted 
inversion parameters of the samples. Indeed, our structural analysis reveals the inversion parameter $\delta$ = 0.76, which is closer to 0.5 than that of 0.84 in Ref.~\onlinecite{petrakovskii_spin-glass_2001}, suggesting stronger disorder in our samples. The disorder would disturb the path of magnetic interactions between Cu$^{2+}$ ions and give rise to randomness of exchange interactions. Once the disorder (exchange randomness) reaches a certain level, it would completely destruct the long-range magnetic order and restrict magnetic correlations within short range, giving rise to a spin-glass ground state. Accordingly, the well-defined spin wave excitations in a magnetically ordered state turn into a continuum-like pattern. Since disorder can drive the system farther away from the magnetically ordered regime, the higher degree of disorder in our samples would lead to lower freezing temperature $T_{\mathrm{f}}$. Taken together, these data provide solid evidence that CuGa$_2$O$_4$ hosts a disorder-induced spin-glass ground state.} It is also worth noting that theoretical studies suggest that sufficiently dilute impurities can induce order, rather than a disordered state, in spinel compounds~\cite{savary_impurity_2011}. In this context, the degree of disorder is increasingly being recognized as a means to modulate the ground state, rather than being merely an "undesirable" perturbation.

Overall, our comprehensive investigations on the spinel oxide \CGO, with significant structural disorder, shed light on the intricate interplay between disorder and the ground state. 
In most cases, structural disorder induces a mimicry of the spin-liquid state rather than a genuine QSL state. In light of this, to find an ideal QSL candidate, it is important to look for materials that are free of structural disorder~\cite{wen_experimental_2019,broholm_quantum_2020,liu_rare-earth_2018,bordelon_field-tunable_2019,dai_spinon_2021}. {\new For instance, target materials should either have magnetic and nonmagnetic ions occupying separate layers or show a notable mismatch in their ionic radii.} Furthermore, materials that have large exchange interactions will also be helpful in realizing the QSL state, as they are more robust to the perturbations and maintain the original QSL state. {\new Along this direction, one can design and explore quantum magnets with shorter and/or more direct exchange paths.} 

\section{Acknowledgements}

We thank Shun-Li Yu, Jian-Xin Li and Qingzhen Huang for stimulating discussions. We acknowledge Shichao Li and Shucui Sun from Quantum Design China for the help in measuring the ac susceptibility. This work was supported by the National Key Projects for Research and Development of China with Grants No.~2021YFA1400400 and No.~2024YFA1409200, National Natural Science Foundation of China with Grants No.~12434005, No.~12225407, No.~12204160, No.~12404173, Postdoctoral Fellowship Program of CPSF under Grant Number BX20240161, China Postdoctoral Science Foundation with
Grant No.~2024M751367, Hubei Provincial Natural Science Foundation of China with Grant No.~2025AFB738, Natural Science Foundation of Jiangsu Province with Grant Nos.~BK20241250, BK20241251, and BK20233001, and Jiangsu Province Excellent Postdoctoral Program with Grant No.~2024ZB021. The work at BNL was supported by the US Department of Energy, Office of Basic Energy Sciences, contract No. DOE-SC0012704.  Access to MACS was provided by the Center for High Resolution Neutron Scattering, a partnership between the National Institute of Standards and Technology and the National Science Foundation under Agreement No. DMR-2010792. We acknowledge the support of the National Institute of Standards and Technology, U.S. Department of Commerce, in providing some of the neutron research facilities used in this work. The identification of x-ray diffractometer (SmartLab SE, Rigaku) and Quantum Design physical property measurement system (PPMS) in this work is solely for the purpose of experimental reproducibility and does not imply endorsement or recommendation by the National Institute of Standards and Technology. A portion of this research at Oak Ridge National Laboratory’s Spallation Neutron Source was sponsored by the U.S. Department of Energy, Office of Basic Energy Sciences with proposal No.~IPTS-24420.


\begin{thebibliography}{74}%
\makeatletter
\providecommand \@ifxundefined [1]{%
 \@ifx{#1\undefined}
}%
\providecommand \@ifnum [1]{%
 \ifnum #1\expandafter \@firstoftwo
 \else \expandafter \@secondoftwo
 \fi
}%
\providecommand \@ifx [1]{%
 \ifx #1\expandafter \@firstoftwo
 \else \expandafter \@secondoftwo
 \fi
}%
\providecommand \natexlab [1]{#1}%
\providecommand \enquote  [1]{``#1''}%
\providecommand \bibnamefont  [1]{#1}%
\providecommand \bibfnamefont [1]{#1}%
\providecommand \citenamefont [1]{#1}%
\providecommand \href@noop [0]{\@secondoftwo}%
\providecommand \href [0]{\begingroup \@sanitize@url \@href}%
\providecommand \@href[1]{\@@startlink{#1}\@@href}%
\providecommand \@@href[1]{\endgroup#1\@@endlink}%
\providecommand \@sanitize@url [0]{\catcode `\\12\catcode `\$12\catcode
  `\&12\catcode `\#12\catcode `\^12\catcode `\_12\catcode `\%12\relax}%
\providecommand \@@startlink[1]{}%
\providecommand \@@endlink[0]{}%
\providecommand \url  [0]{\begingroup\@sanitize@url \@url }%
\providecommand \@url [1]{\endgroup\@href {#1}{\urlprefix }}%
\providecommand \urlprefix  [0]{URL }%
\providecommand \Eprint [0]{\href }%
\providecommand \doibase [0]{http://dx.doi.org/}%
\providecommand \selectlanguage [0]{\@gobble}%
\providecommand \bibinfo  [0]{\@secondoftwo}%
\providecommand \bibfield  [0]{\@secondoftwo}%
\providecommand \translation [1]{[#1]}%
\providecommand \BibitemOpen [0]{}%
\providecommand \bibitemStop [0]{}%
\providecommand \bibitemNoStop [0]{.\EOS\space}%
\providecommand \EOS [0]{\spacefactor3000\relax}%
\providecommand \BibitemShut  [1]{\csname bibitem#1\endcsname}%
\let\auto@bib@innerbib\@empty
\bibitem [{\citenamefont {Wannier}(1950)}]{wannier_antiferromagnetism_1950}%
  \BibitemOpen
  \bibfield  {author} {\bibinfo {author} {\bibfnamefont {G.~H.}\ \bibnamefont
  {Wannier}},\ }\bibfield  {title} {\enquote {\bibinfo {title}
  {Antiferromagnetism. {{The Triangular Ising Net}}},}\ }\href {\doibase
  10.1103/PhysRev.79.357} {\bibfield  {journal} {\bibinfo  {journal} {Phys.
  Rev.}\ }\textbf {\bibinfo {volume} {79}},\ \bibinfo {pages} {357} (\bibinfo
  {year} {1950})}\BibitemShut {NoStop}%
\bibitem [{\citenamefont {Gardner}\ \emph {et~al.}(2010)\citenamefont
  {Gardner}, \citenamefont {Gingras},\ and\ \citenamefont
  {Greedan}}]{gardner_magnetic_2010}%
  \BibitemOpen
  \bibfield  {author} {\bibinfo {author} {\bibfnamefont {Jason~S.}\
  \bibnamefont {Gardner}}, \bibinfo {author} {\bibfnamefont {Michel J.~P.}\
  \bibnamefont {Gingras}}, \ and\ \bibinfo {author} {\bibfnamefont {John~E.}\
  \bibnamefont {Greedan}},\ }\bibfield  {title} {\enquote {\bibinfo {title}
  {Magnetic pyrochlore oxides},}\ }\href {\doibase 10.1103/RevModPhys.82.53}
  {\bibfield  {journal} {\bibinfo  {journal} {Rev. Mod. Phys.}\ }\textbf
  {\bibinfo {volume} {82}},\ \bibinfo {pages} {53} (\bibinfo {year}
  {2010})}\BibitemShut {NoStop}%
\bibitem [{\citenamefont {Balents}(2010)}]{balents_spin_2010}%
  \BibitemOpen
  \bibfield  {author} {\bibinfo {author} {\bibfnamefont {Leon}\ \bibnamefont
  {Balents}},\ }\bibfield  {title} {\enquote {\bibinfo {title} {Spin liquids in
  frustrated magnets},}\ }\href {\doibase 10.1038/nature08917} {\bibfield
  {journal} {\bibinfo  {journal} {Nature}\ }\textbf {\bibinfo {volume} {464}},\
  \bibinfo {pages} {199} (\bibinfo {year} {2010})}\BibitemShut {NoStop}%
\bibitem [{\citenamefont {Anderson}(1973)}]{anderson_resonating_1973}%
  \BibitemOpen
  \bibfield  {author} {\bibinfo {author} {\bibfnamefont {P.W.}\ \bibnamefont
  {Anderson}},\ }\bibfield  {title} {\enquote {\bibinfo {title} {Resonating
  valence bonds: {{A}} new kind of insulator?}}\ }\href {\doibase
  10.1016/0025-5408(73)90167-0} {\bibfield  {journal} {\bibinfo  {journal}
  {Mater. Res. Bull.}\ }\textbf {\bibinfo {volume} {8}},\ \bibinfo {pages}
  {153} (\bibinfo {year} {1973})}\BibitemShut {NoStop}%
\bibitem [{\citenamefont {Zhou}\ \emph {et~al.}(2017)\citenamefont {Zhou},
  \citenamefont {Kanoda},\ and\ \citenamefont {Ng}}]{zhou_quantum_2017}%
  \BibitemOpen
  \bibfield  {author} {\bibinfo {author} {\bibfnamefont {Yi}~\bibnamefont
  {Zhou}}, \bibinfo {author} {\bibfnamefont {Kazushi}\ \bibnamefont {Kanoda}},
  \ and\ \bibinfo {author} {\bibfnamefont {Tai-Kai}\ \bibnamefont {Ng}},\
  }\bibfield  {title} {\enquote {\bibinfo {title} {Quantum spin liquid
  states},}\ }\href {\doibase 10.1103/RevModPhys.89.025003} {\bibfield
  {journal} {\bibinfo  {journal} {Rev. Mod. Phys.}\ }\textbf {\bibinfo {volume}
  {89}},\ \bibinfo {pages} {025003} (\bibinfo {year} {2017})}\BibitemShut
  {NoStop}%
\bibitem [{\citenamefont {Wen}\ \emph {et~al.}(2019)\citenamefont {Wen},
  \citenamefont {Yu}, \citenamefont {Li}, \citenamefont {Yu},\ and\
  \citenamefont {Li}}]{wen_experimental_2019}%
  \BibitemOpen
  \bibfield  {author} {\bibinfo {author} {\bibfnamefont {Jinsheng}\
  \bibnamefont {Wen}}, \bibinfo {author} {\bibfnamefont {Shun-Li}\ \bibnamefont
  {Yu}}, \bibinfo {author} {\bibfnamefont {Shiyan}\ \bibnamefont {Li}},
  \bibinfo {author} {\bibfnamefont {Weiqiang}\ \bibnamefont {Yu}}, \ and\
  \bibinfo {author} {\bibfnamefont {Jian-Xin}\ \bibnamefont {Li}},\ }\bibfield
  {title} {\enquote {\bibinfo {title} {Experimental identification of quantum
  spin liquids},}\ }\href {\doibase 10.1038/s41535-019-0151-6} {\bibfield
  {journal} {\bibinfo  {journal} {npj Quantum Mater.}\ }\textbf {\bibinfo
  {volume} {4}},\ \bibinfo {pages} {12} (\bibinfo {year} {2019})}\BibitemShut
  {NoStop}%
\bibitem [{\citenamefont {Knolle}\ and\ \citenamefont
  {Moessner}(2019)}]{knolle_field_2019}%
  \BibitemOpen
  \bibfield  {author} {\bibinfo {author} {\bibfnamefont {J.}~\bibnamefont
  {Knolle}}\ and\ \bibinfo {author} {\bibfnamefont {R.}~\bibnamefont
  {Moessner}},\ }\bibfield  {title} {\enquote {\bibinfo {title} {A {{Field
  Guide}} to {{Spin Liquids}}},}\ }\href {\doibase
  10.1146/annurev-conmatphys-031218-013401} {\bibfield  {journal} {\bibinfo
  {journal} {Annu. Rev. Condens. Matter Phys.}\ }\textbf {\bibinfo {volume}
  {10}},\ \bibinfo {pages} {451} (\bibinfo {year} {2019})}\BibitemShut
  {NoStop}%
\bibitem [{\citenamefont {Broholm}\ \emph {et~al.}(2020)\citenamefont
  {Broholm}, \citenamefont {Cava}, \citenamefont {Kivelson}, \citenamefont
  {Nocera}, \citenamefont {Norman},\ and\ \citenamefont
  {Senthil}}]{broholm_quantum_2020}%
  \BibitemOpen
  \bibfield  {author} {\bibinfo {author} {\bibfnamefont {C.}~\bibnamefont
  {Broholm}}, \bibinfo {author} {\bibfnamefont {R.~J.}\ \bibnamefont {Cava}},
  \bibinfo {author} {\bibfnamefont {S.~A.}\ \bibnamefont {Kivelson}}, \bibinfo
  {author} {\bibfnamefont {D.~G.}\ \bibnamefont {Nocera}}, \bibinfo {author}
  {\bibfnamefont {M.~R.}\ \bibnamefont {Norman}}, \ and\ \bibinfo {author}
  {\bibfnamefont {T.}~\bibnamefont {Senthil}},\ }\bibfield  {title} {\enquote
  {\bibinfo {title} {Quantum spin liquids},}\ }\href {\doibase
  10.1126/science.aay0668} {\bibfield  {journal} {\bibinfo  {journal}
  {Science}\ }\textbf {\bibinfo {volume} {367}},\ \bibinfo {pages} {eaay0668}
  (\bibinfo {year} {2020})}\BibitemShut {NoStop}%
\bibitem [{\citenamefont {Han}\ \emph {et~al.}(2012)\citenamefont {Han},
  \citenamefont {Helton}, \citenamefont {Chu}, \citenamefont {Nocera},
  \citenamefont {{Rodriguez-Rivera}}, \citenamefont {Broholm},\ and\
  \citenamefont {Lee}}]{han_fractionalized_2012}%
  \BibitemOpen
  \bibfield  {author} {\bibinfo {author} {\bibfnamefont {Tian-Heng}\
  \bibnamefont {Han}}, \bibinfo {author} {\bibfnamefont {Joel~S.}\ \bibnamefont
  {Helton}}, \bibinfo {author} {\bibfnamefont {Shaoyan}\ \bibnamefont {Chu}},
  \bibinfo {author} {\bibfnamefont {Daniel~G.}\ \bibnamefont {Nocera}},
  \bibinfo {author} {\bibfnamefont {Jose~A.}\ \bibnamefont
  {{Rodriguez-Rivera}}}, \bibinfo {author} {\bibfnamefont {Collin}\
  \bibnamefont {Broholm}}, \ and\ \bibinfo {author} {\bibfnamefont {Young~S.}\
  \bibnamefont {Lee}},\ }\bibfield  {title} {\enquote {\bibinfo {title}
  {Fractionalized excitations in the spin-liquid state of a kagome-lattice
  antiferromagnet},}\ }\href {\doibase 10.1038/nature11659} {\bibfield
  {journal} {\bibinfo  {journal} {Nature}\ }\textbf {\bibinfo {volume} {492}},\
  \bibinfo {pages} {406} (\bibinfo {year} {2012})}\BibitemShut {NoStop}%
\bibitem [{\citenamefont {Paddison}\ \emph {et~al.}(2017)\citenamefont
  {Paddison}, \citenamefont {Daum}, \citenamefont {Dun}, \citenamefont
  {Ehlers}, \citenamefont {Liu}, \citenamefont {Stone}, \citenamefont {Zhou},\
  and\ \citenamefont {Mourigal}}]{paddison_continuous_2017}%
  \BibitemOpen
  \bibfield  {author} {\bibinfo {author} {\bibfnamefont {Joseph A.~M.}\
  \bibnamefont {Paddison}}, \bibinfo {author} {\bibfnamefont {Marcus}\
  \bibnamefont {Daum}}, \bibinfo {author} {\bibfnamefont {Zhiling}\
  \bibnamefont {Dun}}, \bibinfo {author} {\bibfnamefont {Georg}\ \bibnamefont
  {Ehlers}}, \bibinfo {author} {\bibfnamefont {Yaohua}\ \bibnamefont {Liu}},
  \bibinfo {author} {\bibfnamefont {Matthew~B.}\ \bibnamefont {Stone}},
  \bibinfo {author} {\bibfnamefont {Haidong}\ \bibnamefont {Zhou}}, \ and\
  \bibinfo {author} {\bibfnamefont {Martin}\ \bibnamefont {Mourigal}},\
  }\bibfield  {title} {\enquote {\bibinfo {title} {Continuous excitations of
  the triangular-lattice quantum spin liquid {{YbMgGaO}}{\textsubscript{4}}},}\
  }\href {\doibase 10.1038/nphys3971} {\bibfield  {journal} {\bibinfo
  {journal} {Nat. Phys.}\ }\textbf {\bibinfo {volume} {13}},\ \bibinfo {pages}
  {117} (\bibinfo {year} {2017})}\BibitemShut {NoStop}%
\bibitem [{\citenamefont {Kitaev}(2006)}]{kitaev_anyons_2006}%
  \BibitemOpen
  \bibfield  {author} {\bibinfo {author} {\bibfnamefont {Alexei}\ \bibnamefont
  {Kitaev}},\ }\bibfield  {title} {\enquote {\bibinfo {title} {Anyons in an
  exactly solved model and beyond},}\ }\href {\doibase
  10.1016/j.aop.2005.10.005} {\bibfield  {journal} {\bibinfo  {journal} {Ann.
  Phys.}\ }\textbf {\bibinfo {volume} {321}},\ \bibinfo {pages} {2} (\bibinfo
  {year} {2006})}\BibitemShut {NoStop}%
\bibitem [{\citenamefont {Ran}\ \emph {et~al.}(2017)\citenamefont {Ran},
  \citenamefont {Wang}, \citenamefont {Wang}, \citenamefont {Dong},
  \citenamefont {Ren}, \citenamefont {Bao}, \citenamefont {Li}, \citenamefont
  {Ma}, \citenamefont {Gan}, \citenamefont {Zhang}, \citenamefont {Park},
  \citenamefont {Deng}, \citenamefont {Danilkin}, \citenamefont {Yu},
  \citenamefont {Li},\ and\ \citenamefont {Wen}}]{ran_spin-wave_2017}%
  \BibitemOpen
  \bibfield  {author} {\bibinfo {author} {\bibfnamefont {Kejing}\ \bibnamefont
  {Ran}}, \bibinfo {author} {\bibfnamefont {Jinghui}\ \bibnamefont {Wang}},
  \bibinfo {author} {\bibfnamefont {Wei}\ \bibnamefont {Wang}}, \bibinfo
  {author} {\bibfnamefont {Zhao-Yang}\ \bibnamefont {Dong}}, \bibinfo {author}
  {\bibfnamefont {Xiao}\ \bibnamefont {Ren}}, \bibinfo {author} {\bibfnamefont
  {Song}\ \bibnamefont {Bao}}, \bibinfo {author} {\bibfnamefont {Shichao}\
  \bibnamefont {Li}}, \bibinfo {author} {\bibfnamefont {Zhen}\ \bibnamefont
  {Ma}}, \bibinfo {author} {\bibfnamefont {Yuan}\ \bibnamefont {Gan}}, \bibinfo
  {author} {\bibfnamefont {Youtian}\ \bibnamefont {Zhang}}, \bibinfo {author}
  {\bibfnamefont {J.~T.}\ \bibnamefont {Park}}, \bibinfo {author}
  {\bibfnamefont {Guochu}\ \bibnamefont {Deng}}, \bibinfo {author}
  {\bibfnamefont {S.}~\bibnamefont {Danilkin}}, \bibinfo {author}
  {\bibfnamefont {Shun-Li}\ \bibnamefont {Yu}}, \bibinfo {author}
  {\bibfnamefont {Jian-Xin}\ \bibnamefont {Li}}, \ and\ \bibinfo {author}
  {\bibfnamefont {Jinsheng}\ \bibnamefont {Wen}},\ }\bibfield  {title}
  {\enquote {\bibinfo {title} {Spin-{{Wave Excitations Evidencing}} the
  {{Kitaev Interaction}} in {{Single Crystalline}}
  {$\alpha-$}{{RuCl}}{\textsubscript{3}}},}\ }\href {\doibase
  10.1103/PhysRevLett.118.107203} {\bibfield  {journal} {\bibinfo  {journal}
  {Phys. Rev. Lett.}\ }\textbf {\bibinfo {volume} {118}},\ \bibinfo {pages}
  {107203} (\bibinfo {year} {2017})}\BibitemShut {NoStop}%
\bibitem [{\citenamefont {Shangguan}\ \emph {et~al.}(2023)\citenamefont
  {Shangguan}, \citenamefont {Bao}, \citenamefont {Dong}, \citenamefont {Xi},
  \citenamefont {Gao}, \citenamefont {Ma}, \citenamefont {Wang}, \citenamefont
  {Qi}, \citenamefont {Zhang}, \citenamefont {Huang}, \citenamefont {Liao},
  \citenamefont {Zhao}, \citenamefont {Zhang}, \citenamefont {Cheng},
  \citenamefont {Xu}, \citenamefont {Yu}, \citenamefont {Mole}, \citenamefont
  {Murai}, \citenamefont {{Ohira-Kawamura}}, \citenamefont {He}, \citenamefont
  {Hao}, \citenamefont {Yan}, \citenamefont {Song}, \citenamefont {Li},
  \citenamefont {Yu}, \citenamefont {Li},\ and\ \citenamefont
  {Wen}}]{shangguan_one-third_2023}%
  \BibitemOpen
  \bibfield  {author} {\bibinfo {author} {\bibfnamefont {Yanyan}\ \bibnamefont
  {Shangguan}}, \bibinfo {author} {\bibfnamefont {Song}\ \bibnamefont {Bao}},
  \bibinfo {author} {\bibfnamefont {Zhao-Yang}\ \bibnamefont {Dong}}, \bibinfo
  {author} {\bibfnamefont {Ning}\ \bibnamefont {Xi}}, \bibinfo {author}
  {\bibfnamefont {Yi-Peng}\ \bibnamefont {Gao}}, \bibinfo {author}
  {\bibfnamefont {Zhen}\ \bibnamefont {Ma}}, \bibinfo {author} {\bibfnamefont
  {Wei}\ \bibnamefont {Wang}}, \bibinfo {author} {\bibfnamefont {Zhongyuan}\
  \bibnamefont {Qi}}, \bibinfo {author} {\bibfnamefont {Shuai}\ \bibnamefont
  {Zhang}}, \bibinfo {author} {\bibfnamefont {Zhentao}\ \bibnamefont {Huang}},
  \bibinfo {author} {\bibfnamefont {Junbo}\ \bibnamefont {Liao}}, \bibinfo
  {author} {\bibfnamefont {Xiaoxue}\ \bibnamefont {Zhao}}, \bibinfo {author}
  {\bibfnamefont {Bo}~\bibnamefont {Zhang}}, \bibinfo {author} {\bibfnamefont
  {Shufan}\ \bibnamefont {Cheng}}, \bibinfo {author} {\bibfnamefont {Hao}\
  \bibnamefont {Xu}}, \bibinfo {author} {\bibfnamefont {Dehong}\ \bibnamefont
  {Yu}}, \bibinfo {author} {\bibfnamefont {Richard~A.}\ \bibnamefont {Mole}},
  \bibinfo {author} {\bibfnamefont {Naoki}\ \bibnamefont {Murai}}, \bibinfo
  {author} {\bibfnamefont {Seiko}\ \bibnamefont {{Ohira-Kawamura}}}, \bibinfo
  {author} {\bibfnamefont {Lunhua}\ \bibnamefont {He}}, \bibinfo {author}
  {\bibfnamefont {Jiazheng}\ \bibnamefont {Hao}}, \bibinfo {author}
  {\bibfnamefont {Qing-Bo}\ \bibnamefont {Yan}}, \bibinfo {author}
  {\bibfnamefont {Fengqi}\ \bibnamefont {Song}}, \bibinfo {author}
  {\bibfnamefont {Wei}\ \bibnamefont {Li}}, \bibinfo {author} {\bibfnamefont
  {Shun-Li}\ \bibnamefont {Yu}}, \bibinfo {author} {\bibfnamefont {Jian-Xin}\
  \bibnamefont {Li}}, \ and\ \bibinfo {author} {\bibfnamefont {Jinsheng}\
  \bibnamefont {Wen}},\ }\bibfield  {title} {\enquote {\bibinfo {title} {A
  one-third magnetization plateau phase as evidence for the {{Kitaev}}
  interaction in a honeycomb-lattice antiferromagnet},}\ }\href {\doibase
  10.1038/s41567-023-02212-2} {\bibfield  {journal} {\bibinfo  {journal} {Nat.
  Phys.}\ }\textbf {\bibinfo {volume} {19}},\ \bibinfo {pages} {1883} (\bibinfo
  {year} {2023})}\BibitemShut {NoStop}%
\bibitem [{\citenamefont {Fiorani}\ \emph {et~al.}(1985)\citenamefont
  {Fiorani}, \citenamefont {Dormann}, \citenamefont {Tholence},\ and\
  \citenamefont {Soubeyroux}}]{fiorani_antiferromagnetic_1985}%
  \BibitemOpen
  \bibfield  {author} {\bibinfo {author} {\bibfnamefont {D}~\bibnamefont
  {Fiorani}}, \bibinfo {author} {\bibfnamefont {J~L}\ \bibnamefont {Dormann}},
  \bibinfo {author} {\bibfnamefont {J~L}\ \bibnamefont {Tholence}}, \ and\
  \bibinfo {author} {\bibfnamefont {J~L}\ \bibnamefont {Soubeyroux}},\
  }\bibfield  {title} {\enquote {\bibinfo {title} {From the antiferromagnetic
  regime to the spin-glass state in the frustrated spinel system
  {{ZnCr}}{\textsubscript{2x}}{{Ga}}{\textsubscript{2-2x}}{{O}}{\textsubscript{4}}},}\
  }\href {\doibase 10.1088/0022-3719/18/15/014} {\bibfield  {journal} {\bibinfo
   {journal} {J. Phys. C: Solid State Phys.}\ }\textbf {\bibinfo {volume}
  {18}},\ \bibinfo {pages} {3053} (\bibinfo {year} {1985})}\BibitemShut
  {NoStop}%
\bibitem [{\citenamefont {Lee}\ \emph {et~al.}(2002)\citenamefont {Lee},
  \citenamefont {Broholm}, \citenamefont {Ratcliff}, \citenamefont
  {Gasparovic}, \citenamefont {Huang}, \citenamefont {Kim},\ and\ \citenamefont
  {Cheong}}]{lee_emergent_2002}%
  \BibitemOpen
  \bibfield  {author} {\bibinfo {author} {\bibfnamefont {S.-H.}\ \bibnamefont
  {Lee}}, \bibinfo {author} {\bibfnamefont {C.}~\bibnamefont {Broholm}},
  \bibinfo {author} {\bibfnamefont {W.}~\bibnamefont {Ratcliff}}, \bibinfo
  {author} {\bibfnamefont {G.}~\bibnamefont {Gasparovic}}, \bibinfo {author}
  {\bibfnamefont {Q.}~\bibnamefont {Huang}}, \bibinfo {author} {\bibfnamefont
  {T.~H.}\ \bibnamefont {Kim}}, \ and\ \bibinfo {author} {\bibfnamefont
  {S.-W.}\ \bibnamefont {Cheong}},\ }\bibfield  {title} {\enquote {\bibinfo
  {title} {Emergent excitations in a geometrically frustrated magnet},}\ }\href
  {\doibase 10.1038/nature00964} {\bibfield  {journal} {\bibinfo  {journal}
  {Nature}\ }\textbf {\bibinfo {volume} {418}},\ \bibinfo {pages} {856}
  (\bibinfo {year} {2002})}\BibitemShut {NoStop}%
\bibitem [{\citenamefont {Bergman}\ \emph {et~al.}(2007)\citenamefont
  {Bergman}, \citenamefont {Alicea}, \citenamefont {Gull}, \citenamefont
  {Trebst},\ and\ \citenamefont {Balents}}]{bergman_order-by-disorder_2007}%
  \BibitemOpen
  \bibfield  {author} {\bibinfo {author} {\bibfnamefont {Doron}\ \bibnamefont
  {Bergman}}, \bibinfo {author} {\bibfnamefont {Jason}\ \bibnamefont {Alicea}},
  \bibinfo {author} {\bibfnamefont {Emanuel}\ \bibnamefont {Gull}}, \bibinfo
  {author} {\bibfnamefont {Simon}\ \bibnamefont {Trebst}}, \ and\ \bibinfo
  {author} {\bibfnamefont {Leon}\ \bibnamefont {Balents}},\ }\bibfield  {title}
  {\enquote {\bibinfo {title} {Order-by-disorder and spiral spin-liquid in
  frustrated diamond-lattice antiferromagnets},}\ }\href {\doibase
  10.1038/nphys622} {\bibfield  {journal} {\bibinfo  {journal} {Nat. Phys.}\
  }\textbf {\bibinfo {volume} {3}},\ \bibinfo {pages} {487} (\bibinfo {year}
  {2007})}\BibitemShut {NoStop}%
\bibitem [{\citenamefont {Zaharko}\ \emph {et~al.}(2010)\citenamefont
  {Zaharko}, \citenamefont {Cervellino}, \citenamefont {Tsurkan}, \citenamefont
  {Christensen},\ and\ \citenamefont {Loidl}}]{zaharko_evolution_2010}%
  \BibitemOpen
  \bibfield  {author} {\bibinfo {author} {\bibfnamefont {O.}~\bibnamefont
  {Zaharko}}, \bibinfo {author} {\bibfnamefont {A.}~\bibnamefont {Cervellino}},
  \bibinfo {author} {\bibfnamefont {V.}~\bibnamefont {Tsurkan}}, \bibinfo
  {author} {\bibfnamefont {N.~B.}\ \bibnamefont {Christensen}}, \ and\ \bibinfo
  {author} {\bibfnamefont {A.}~\bibnamefont {Loidl}},\ }\bibfield  {title}
  {\enquote {\bibinfo {title} {Evolution of magnetic states in frustrated
  diamond lattice antiferromagnetic
  {{Co}}({{Al}}{\textsubscript{1-x}}{{Co}}{\textsubscript{x}}){\textsubscript{2}}{{O}}{\textsubscript{4}}
  spinels},}\ }\href {\doibase 10.1103/PhysRevB.81.064416} {\bibfield
  {journal} {\bibinfo  {journal} {Phys. Rev. B}\ }\textbf {\bibinfo {volume}
  {81}},\ \bibinfo {pages} {064416} (\bibinfo {year} {2010})}\BibitemShut
  {NoStop}%
\bibitem [{\citenamefont {Savary}\ \emph {et~al.}(2011)\citenamefont {Savary},
  \citenamefont {Gull}, \citenamefont {Trebst}, \citenamefont {Alicea},
  \citenamefont {Bergman},\ and\ \citenamefont
  {Balents}}]{savary_impurity_2011}%
  \BibitemOpen
  \bibfield  {author} {\bibinfo {author} {\bibfnamefont {Lucile}\ \bibnamefont
  {Savary}}, \bibinfo {author} {\bibfnamefont {Emanuel}\ \bibnamefont {Gull}},
  \bibinfo {author} {\bibfnamefont {Simon}\ \bibnamefont {Trebst}}, \bibinfo
  {author} {\bibfnamefont {Jason}\ \bibnamefont {Alicea}}, \bibinfo {author}
  {\bibfnamefont {Doron}\ \bibnamefont {Bergman}}, \ and\ \bibinfo {author}
  {\bibfnamefont {Leon}\ \bibnamefont {Balents}},\ }\bibfield  {title}
  {\enquote {\bibinfo {title} {Impurity effects in highly frustrated
  diamond-lattice antiferromagnets},}\ }\href {\doibase
  10.1103/PhysRevB.84.064438} {\bibfield  {journal} {\bibinfo  {journal} {Phys.
  Rev. B}\ }\textbf {\bibinfo {volume} {84}},\ \bibinfo {pages} {064438}
  (\bibinfo {year} {2011})}\BibitemShut {NoStop}%
\bibitem [{\citenamefont {Henley}(1989)}]{henley_ordering_1989}%
  \BibitemOpen
  \bibfield  {author} {\bibinfo {author} {\bibfnamefont {Christopher~L.}\
  \bibnamefont {Henley}},\ }\bibfield  {title} {\enquote {\bibinfo {title}
  {Ordering due to disorder in a frustrated vector antiferromagnet},}\ }\href
  {\doibase 10.1103/PhysRevLett.62.2056} {\bibfield  {journal} {\bibinfo
  {journal} {Phys. Rev. Lett.}\ }\textbf {\bibinfo {volume} {62}},\ \bibinfo
  {pages} {2056} (\bibinfo {year} {1989})}\BibitemShut {NoStop}%
\bibitem [{\citenamefont {Tristan}\ \emph {et~al.}(2005)\citenamefont
  {Tristan}, \citenamefont {Hemberger}, \citenamefont {Krimmel}, \citenamefont
  {Krug Von~Nidda}, \citenamefont {Tsurkan},\ and\ \citenamefont
  {Loidl}}]{tristan_geometric_2005}%
  \BibitemOpen
  \bibfield  {author} {\bibinfo {author} {\bibfnamefont {N.}~\bibnamefont
  {Tristan}}, \bibinfo {author} {\bibfnamefont {J.}~\bibnamefont {Hemberger}},
  \bibinfo {author} {\bibfnamefont {A.}~\bibnamefont {Krimmel}}, \bibinfo
  {author} {\bibfnamefont {H-A.}\ \bibnamefont {Krug Von~Nidda}}, \bibinfo
  {author} {\bibfnamefont {V.}~\bibnamefont {Tsurkan}}, \ and\ \bibinfo
  {author} {\bibfnamefont {A.}~\bibnamefont {Loidl}},\ }\bibfield  {title}
  {\enquote {\bibinfo {title} {Geometric frustration in the cubic spinels
  {{MAl}}{\textsubscript{2}}{{O}}{\textsubscript{4}} ({{M}} = {{Co}}, {{Fe}},
  and {{Mn}})},}\ }\href {\doibase 10.1103/PhysRevB.72.174404} {\bibfield
  {journal} {\bibinfo  {journal} {Phys. Rev. B}\ }\textbf {\bibinfo {volume}
  {72}},\ \bibinfo {pages} {174404} (\bibinfo {year} {2005})}\BibitemShut
  {NoStop}%
\bibitem [{\citenamefont {Bernier}\ \emph {et~al.}(2008)\citenamefont
  {Bernier}, \citenamefont {Lawler},\ and\ \citenamefont
  {Kim}}]{bernier_quantum_2008}%
  \BibitemOpen
  \bibfield  {author} {\bibinfo {author} {\bibfnamefont {Jean-S{\'e}bastien}\
  \bibnamefont {Bernier}}, \bibinfo {author} {\bibfnamefont {Michael~J.}\
  \bibnamefont {Lawler}}, \ and\ \bibinfo {author} {\bibfnamefont {Yong~Baek}\
  \bibnamefont {Kim}},\ }\bibfield  {title} {\enquote {\bibinfo {title}
  {Quantum {{Order}} by {{Disorder}} in {{Frustrated Diamond Lattice
  Antiferromagnets}}},}\ }\href {\doibase 10.1103/PhysRevLett.101.047201}
  {\bibfield  {journal} {\bibinfo  {journal} {Phys. Rev. Lett.}\ }\textbf
  {\bibinfo {volume} {101}},\ \bibinfo {pages} {047201} (\bibinfo {year}
  {2008})}\BibitemShut {NoStop}%
\bibitem [{\citenamefont {MacDougall}\ \emph {et~al.}(2011)\citenamefont
  {MacDougall}, \citenamefont {Gout}, \citenamefont {Zarestky}, \citenamefont
  {Ehlers}, \citenamefont {Podlesnyak}, \citenamefont {McGuire}, \citenamefont
  {Mandrus},\ and\ \citenamefont {Nagler}}]{macdougall_kinetically_2011}%
  \BibitemOpen
  \bibfield  {author} {\bibinfo {author} {\bibfnamefont {Gregory~J.}\
  \bibnamefont {MacDougall}}, \bibinfo {author} {\bibfnamefont {Delphine}\
  \bibnamefont {Gout}}, \bibinfo {author} {\bibfnamefont {Jerel~L.}\
  \bibnamefont {Zarestky}}, \bibinfo {author} {\bibfnamefont {Georg}\
  \bibnamefont {Ehlers}}, \bibinfo {author} {\bibfnamefont {Andrey}\
  \bibnamefont {Podlesnyak}}, \bibinfo {author} {\bibfnamefont {Michael~A.}\
  \bibnamefont {McGuire}}, \bibinfo {author} {\bibfnamefont {David}\
  \bibnamefont {Mandrus}}, \ and\ \bibinfo {author} {\bibfnamefont
  {Stephen~E.}\ \bibnamefont {Nagler}},\ }\bibfield  {title} {\enquote
  {\bibinfo {title} {Kinetically inhibited order in a diamond-lattice
  antiferromagnet},}\ }\href {\doibase 10.1073/pnas.1107861108} {\bibfield
  {journal} {\bibinfo  {journal} {Proc. Natl. Acad. Sci.}\ }\textbf {\bibinfo
  {volume} {108}},\ \bibinfo {pages} {15693} (\bibinfo {year}
  {2011})}\BibitemShut {NoStop}%
\bibitem [{\citenamefont {Zaharko}\ \emph {et~al.}(2011)\citenamefont
  {Zaharko}, \citenamefont {Christensen}, \citenamefont {Cervellino},
  \citenamefont {Tsurkan}, \citenamefont {Maljuk}, \citenamefont {Stuhr},
  \citenamefont {Niedermayer}, \citenamefont {Yokaichiya}, \citenamefont
  {Argyriou}, \citenamefont {Boehm},\ and\ \citenamefont
  {Loidl}}]{zaharko_spin_2011}%
  \BibitemOpen
  \bibfield  {author} {\bibinfo {author} {\bibfnamefont {O.}~\bibnamefont
  {Zaharko}}, \bibinfo {author} {\bibfnamefont {N.~B.}\ \bibnamefont
  {Christensen}}, \bibinfo {author} {\bibfnamefont {A.}~\bibnamefont
  {Cervellino}}, \bibinfo {author} {\bibfnamefont {V.}~\bibnamefont {Tsurkan}},
  \bibinfo {author} {\bibfnamefont {A.}~\bibnamefont {Maljuk}}, \bibinfo
  {author} {\bibfnamefont {U.}~\bibnamefont {Stuhr}}, \bibinfo {author}
  {\bibfnamefont {C.}~\bibnamefont {Niedermayer}}, \bibinfo {author}
  {\bibfnamefont {F.}~\bibnamefont {Yokaichiya}}, \bibinfo {author}
  {\bibfnamefont {D.~N.}\ \bibnamefont {Argyriou}}, \bibinfo {author}
  {\bibfnamefont {M.}~\bibnamefont {Boehm}}, \ and\ \bibinfo {author}
  {\bibfnamefont {A.}~\bibnamefont {Loidl}},\ }\bibfield  {title} {\enquote
  {\bibinfo {title} {Spin liquid in a single crystal of the frustrated diamond
  lattice antiferromagnet
  {{CoAl}}{\textsubscript{2}}{{O}}{\textsubscript{4}}},}\ }\href {\doibase
  10.1103/PhysRevB.84.094403} {\bibfield  {journal} {\bibinfo  {journal} {Phys.
  Rev. B}\ }\textbf {\bibinfo {volume} {84}},\ \bibinfo {pages} {094403}
  (\bibinfo {year} {2011})}\BibitemShut {NoStop}%
\bibitem [{\citenamefont {Iakovleva}\ \emph {et~al.}(2015)\citenamefont
  {Iakovleva}, \citenamefont {Vavilova}, \citenamefont {Grafe}, \citenamefont
  {Zimmermann}, \citenamefont {Alfonsov}, \citenamefont {Luetkens},
  \citenamefont {Klauss}, \citenamefont {Maljuk}, \citenamefont {Wurmehl},
  \citenamefont {B{\"u}chner},\ and\ \citenamefont
  {Kataev}}]{iakovleva_ground_2015}%
  \BibitemOpen
  \bibfield  {author} {\bibinfo {author} {\bibfnamefont {M.}~\bibnamefont
  {Iakovleva}}, \bibinfo {author} {\bibfnamefont {E.}~\bibnamefont {Vavilova}},
  \bibinfo {author} {\bibfnamefont {H.-J.}\ \bibnamefont {Grafe}}, \bibinfo
  {author} {\bibfnamefont {S.}~\bibnamefont {Zimmermann}}, \bibinfo {author}
  {\bibfnamefont {A.}~\bibnamefont {Alfonsov}}, \bibinfo {author}
  {\bibfnamefont {H.}~\bibnamefont {Luetkens}}, \bibinfo {author}
  {\bibfnamefont {H.-H.}\ \bibnamefont {Klauss}}, \bibinfo {author}
  {\bibfnamefont {A.}~\bibnamefont {Maljuk}}, \bibinfo {author} {\bibfnamefont
  {S.}~\bibnamefont {Wurmehl}}, \bibinfo {author} {\bibfnamefont
  {B.}~\bibnamefont {B{\"u}chner}}, \ and\ \bibinfo {author} {\bibfnamefont
  {V.}~\bibnamefont {Kataev}},\ }\bibfield  {title} {\enquote {\bibinfo {title}
  {Ground state and low-energy magnetic dynamics in the frustrated magnet
  {{CoAl}}{\textsubscript{2}}{{O}}{\textsubscript{4}} as revealed by local spin
  probes},}\ }\href {\doibase 10.1103/PhysRevB.91.144419} {\bibfield  {journal}
  {\bibinfo  {journal} {Phys. Rev. B}\ }\textbf {\bibinfo {volume} {91}},\
  \bibinfo {pages} {144419} (\bibinfo {year} {2015})}\BibitemShut {NoStop}%
\bibitem [{\citenamefont {Gao}\ \emph {et~al.}(2017)\citenamefont {Gao},
  \citenamefont {Zaharko}, \citenamefont {Tsurkan}, \citenamefont {Su},
  \citenamefont {White}, \citenamefont {Tucker}, \citenamefont {Roessli},
  \citenamefont {Bourdarot}, \citenamefont {Sibille}, \citenamefont
  {Chernyshov}, \citenamefont {Fennell}, \citenamefont {Loidl},\ and\
  \citenamefont {R{\"u}egg}}]{gao_spiral_2017}%
  \BibitemOpen
  \bibfield  {author} {\bibinfo {author} {\bibfnamefont {Shang}\ \bibnamefont
  {Gao}}, \bibinfo {author} {\bibfnamefont {Oksana}\ \bibnamefont {Zaharko}},
  \bibinfo {author} {\bibfnamefont {Vladimir}\ \bibnamefont {Tsurkan}},
  \bibinfo {author} {\bibfnamefont {Yixi}\ \bibnamefont {Su}}, \bibinfo
  {author} {\bibfnamefont {Jonathan~S.}\ \bibnamefont {White}}, \bibinfo
  {author} {\bibfnamefont {Gregory~S.}\ \bibnamefont {Tucker}}, \bibinfo
  {author} {\bibfnamefont {Bertrand}\ \bibnamefont {Roessli}}, \bibinfo
  {author} {\bibfnamefont {Frederic}\ \bibnamefont {Bourdarot}}, \bibinfo
  {author} {\bibfnamefont {Romain}\ \bibnamefont {Sibille}}, \bibinfo {author}
  {\bibfnamefont {Dmitry}\ \bibnamefont {Chernyshov}}, \bibinfo {author}
  {\bibfnamefont {Tom}\ \bibnamefont {Fennell}}, \bibinfo {author}
  {\bibfnamefont {Alois}\ \bibnamefont {Loidl}}, \ and\ \bibinfo {author}
  {\bibfnamefont {Christian}\ \bibnamefont {R{\"u}egg}},\ }\bibfield  {title}
  {\enquote {\bibinfo {title} {Spiral spin-liquid and the emergence of a
  vortex-like state in {{MnSc}}{\textsubscript{2}}{{S}}{\textsubscript{4}}},}\
  }\href {\doibase 10.1038/nphys3914} {\bibfield  {journal} {\bibinfo
  {journal} {Nat. Phys.}\ }\textbf {\bibinfo {volume} {13}},\ \bibinfo {pages}
  {157} (\bibinfo {year} {2017})}\BibitemShut {NoStop}%
\bibitem [{\citenamefont {Nirmala}\ \emph {et~al.}(2017)\citenamefont
  {Nirmala}, \citenamefont {Jang}, \citenamefont {Sim}, \citenamefont {Cho},
  \citenamefont {Lee}, \citenamefont {Yang}, \citenamefont {Lee}, \citenamefont
  {Ibberson}, \citenamefont {Kakurai}, \citenamefont {Matsuda}, \citenamefont
  {Cheong}, \citenamefont {Gapontsev}, \citenamefont {Streltsov},\ and\
  \citenamefont {Park}}]{nirmala_spin_2017}%
  \BibitemOpen
  \bibfield  {author} {\bibinfo {author} {\bibfnamefont {R}~\bibnamefont
  {Nirmala}}, \bibinfo {author} {\bibfnamefont {Kwang-Hyun}\ \bibnamefont
  {Jang}}, \bibinfo {author} {\bibfnamefont {Hasung}\ \bibnamefont {Sim}},
  \bibinfo {author} {\bibfnamefont {Hwanbeom}\ \bibnamefont {Cho}}, \bibinfo
  {author} {\bibfnamefont {Junghwan}\ \bibnamefont {Lee}}, \bibinfo {author}
  {\bibfnamefont {Nam-Geun}\ \bibnamefont {Yang}}, \bibinfo {author}
  {\bibfnamefont {Seongsu}\ \bibnamefont {Lee}}, \bibinfo {author}
  {\bibfnamefont {R~M}\ \bibnamefont {Ibberson}}, \bibinfo {author}
  {\bibfnamefont {K}~\bibnamefont {Kakurai}}, \bibinfo {author} {\bibfnamefont
  {M}~\bibnamefont {Matsuda}}, \bibinfo {author} {\bibfnamefont {S-W}\
  \bibnamefont {Cheong}}, \bibinfo {author} {\bibfnamefont {V~V}\ \bibnamefont
  {Gapontsev}}, \bibinfo {author} {\bibfnamefont {S~V}\ \bibnamefont
  {Streltsov}}, \ and\ \bibinfo {author} {\bibfnamefont {Je-Geun}\ \bibnamefont
  {Park}},\ }\bibfield  {title} {\enquote {\bibinfo {title} {Spin glass
  behavior in frustrated quantum spin system
  {{CuAl}}{\textsubscript{2}}{{O}}{\textsubscript{4}} with a possible orbital
  liquid state},}\ }\href {\doibase 10.1088/1361-648X/aa5c72} {\bibfield
  {journal} {\bibinfo  {journal} {J. Phys.: Condens. Matter}\ }\textbf
  {\bibinfo {volume} {29}},\ \bibinfo {pages} {13LT01} (\bibinfo {year}
  {2017})}\BibitemShut {NoStop}%
\bibitem [{\citenamefont {Nikolaev}\ \emph {et~al.}(2018)\citenamefont
  {Nikolaev}, \citenamefont {Solovyev}, \citenamefont {Ignatenko},
  \citenamefont {Irkhin},\ and\ \citenamefont
  {Streltsov}}]{nikolaev_realization_2018}%
  \BibitemOpen
  \bibfield  {author} {\bibinfo {author} {\bibfnamefont {S.~A.}\ \bibnamefont
  {Nikolaev}}, \bibinfo {author} {\bibfnamefont {I.~V.}\ \bibnamefont
  {Solovyev}}, \bibinfo {author} {\bibfnamefont {A.~N.}\ \bibnamefont
  {Ignatenko}}, \bibinfo {author} {\bibfnamefont {V.~{\relax Yu}.}\
  \bibnamefont {Irkhin}}, \ and\ \bibinfo {author} {\bibfnamefont {S.~V.}\
  \bibnamefont {Streltsov}},\ }\bibfield  {title} {\enquote {\bibinfo {title}
  {Realization of the anisotropic compass model on the diamond lattice of
  {{Cu}}{\textsuperscript{2+}} in
  {{CuAl}}{\textsubscript{2}}{{O}}{\textsubscript{4}}},}\ }\href {\doibase
  10.1103/PhysRevB.98.201106} {\bibfield  {journal} {\bibinfo  {journal} {Phys.
  Rev. B}\ }\textbf {\bibinfo {volume} {98}},\ \bibinfo {pages} {201106}
  (\bibinfo {year} {2018})}\BibitemShut {NoStop}%
\bibitem [{\citenamefont {Huang}\ \emph {et~al.}(2022)\citenamefont {Huang},
  \citenamefont {Singh}, \citenamefont {Wu}, \citenamefont {Xie}, \citenamefont
  {Okamoto}, \citenamefont {Belik}, \citenamefont {Kurmaev}, \citenamefont
  {Fujimori}, \citenamefont {Chen}, \citenamefont {Streltsov},\ and\
  \citenamefont {Huang}}]{huang_resonant_2022}%
  \BibitemOpen
  \bibfield  {author} {\bibinfo {author} {\bibfnamefont {H.~Y.}\ \bibnamefont
  {Huang}}, \bibinfo {author} {\bibfnamefont {A.}~\bibnamefont {Singh}},
  \bibinfo {author} {\bibfnamefont {C.~I.}\ \bibnamefont {Wu}}, \bibinfo
  {author} {\bibfnamefont {J.~D.}\ \bibnamefont {Xie}}, \bibinfo {author}
  {\bibfnamefont {J.}~\bibnamefont {Okamoto}}, \bibinfo {author} {\bibfnamefont
  {A.~A.}\ \bibnamefont {Belik}}, \bibinfo {author} {\bibfnamefont
  {E.}~\bibnamefont {Kurmaev}}, \bibinfo {author} {\bibfnamefont
  {A.}~\bibnamefont {Fujimori}}, \bibinfo {author} {\bibfnamefont {C.~T.}\
  \bibnamefont {Chen}}, \bibinfo {author} {\bibfnamefont {S.~V.}\ \bibnamefont
  {Streltsov}}, \ and\ \bibinfo {author} {\bibfnamefont {D.~J.}\ \bibnamefont
  {Huang}},\ }\bibfield  {title} {\enquote {\bibinfo {title} {Resonant
  inelastic {{X-ray}} scattering as a probe of {{Jeff}} = 1/2 state in 3d
  transition-metal oxide},}\ }\href {\doibase 10.1038/s41535-022-00430-0}
  {\bibfield  {journal} {\bibinfo  {journal} {npj Quantum Mater.}\ }\textbf
  {\bibinfo {volume} {7}},\ \bibinfo {pages} {1--6} (\bibinfo {year}
  {2022})}\BibitemShut {NoStop}%
\bibitem [{\citenamefont {O'Neill}\ and\ \citenamefont
  {Navrotsky}(1983)}]{oneill_simple_1983}%
  \BibitemOpen
  \bibfield  {author} {\bibinfo {author} {\bibfnamefont {Hugh St.~C.}\
  \bibnamefont {O'Neill}}\ and\ \bibinfo {author} {\bibfnamefont {Alexandra}\
  \bibnamefont {Navrotsky}},\ }\bibfield  {title} {\enquote {\bibinfo {title}
  {Simple spinels; crystallographic parameters, cation radii, lattice energies,
  and cation distribution},}\ }\href@noop {} {\bibfield  {journal} {\bibinfo
  {journal} {Am. Mineral.}\ }\textbf {\bibinfo {volume} {68}},\ \bibinfo
  {pages} {181} (\bibinfo {year} {1983})}\BibitemShut {NoStop}%
\bibitem [{\citenamefont {Fregola}\ \emph {et~al.}(2012)\citenamefont
  {Fregola}, \citenamefont {Bosi}, \citenamefont {Skogby},\ and\ \citenamefont
  {H{\aa}lenius}}]{fregola_cation_2012}%
  \BibitemOpen
  \bibfield  {author} {\bibinfo {author} {\bibfnamefont {Rosa~Anna}\
  \bibnamefont {Fregola}}, \bibinfo {author} {\bibfnamefont {Ferdinando}\
  \bibnamefont {Bosi}}, \bibinfo {author} {\bibfnamefont {Henrik}\ \bibnamefont
  {Skogby}}, \ and\ \bibinfo {author} {\bibfnamefont {Ulf}\ \bibnamefont
  {H{\aa}lenius}},\ }\bibfield  {title} {\enquote {\bibinfo {title} {Cation
  ordering over short-range and long-range scales in the
  {{MgAl}}{\textsubscript{2}}{{O}}{\textsubscript{4}}-{{CuAl}}{\textsubscript{2}}{{O}}{\textsubscript{4}}
  series},}\ }\href {\doibase 10.2138/am.2012.4177} {\bibfield  {journal}
  {\bibinfo  {journal} {Am. Mineral.}\ }\textbf {\bibinfo {volume} {97}},\
  \bibinfo {pages} {1821--1827} (\bibinfo {year} {2012})}\BibitemShut {NoStop}%
\bibitem [{\citenamefont {Cho}\ \emph {et~al.}(2020)\citenamefont {Cho},
  \citenamefont {Nirmala}, \citenamefont {Jeong}, \citenamefont {Baker},
  \citenamefont {Takeda}, \citenamefont {Mera}, \citenamefont {Blundell},
  \citenamefont {Takigawa}, \citenamefont {Adroja},\ and\ \citenamefont
  {Park}}]{cho_dynamic_2020}%
  \BibitemOpen
  \bibfield  {author} {\bibinfo {author} {\bibfnamefont {Hwanbeom}\
  \bibnamefont {Cho}}, \bibinfo {author} {\bibfnamefont {R.}~\bibnamefont
  {Nirmala}}, \bibinfo {author} {\bibfnamefont {Jaehong}\ \bibnamefont
  {Jeong}}, \bibinfo {author} {\bibfnamefont {Peter~J.}\ \bibnamefont {Baker}},
  \bibinfo {author} {\bibfnamefont {Hikaru}\ \bibnamefont {Takeda}}, \bibinfo
  {author} {\bibfnamefont {Nobuyoshi}\ \bibnamefont {Mera}}, \bibinfo {author}
  {\bibfnamefont {Stephen~J.}\ \bibnamefont {Blundell}}, \bibinfo {author}
  {\bibfnamefont {Masashi}\ \bibnamefont {Takigawa}}, \bibinfo {author}
  {\bibfnamefont {D.~T.}\ \bibnamefont {Adroja}}, \ and\ \bibinfo {author}
  {\bibfnamefont {Je-Geun}\ \bibnamefont {Park}},\ }\bibfield  {title}
  {\enquote {\bibinfo {title} {Dynamic spin fluctuations in the frustrated
  {$A$}-site spinel {$\mathrm{Cu}{\mathrm{Al}}_{2}{\mathrm{O}}_{4}$}},}\ }\href
  {\doibase 10.1103/PhysRevB.102.014439} {\bibfield  {journal} {\bibinfo
  {journal} {Phys. Rev. B}\ }\textbf {\bibinfo {volume} {102}},\ \bibinfo
  {pages} {14439} (\bibinfo {year} {2020})}\BibitemShut {NoStop}%
\bibitem [{\citenamefont {Furukawa}\ \emph {et~al.}(2015)\citenamefont
  {Furukawa}, \citenamefont {Miyagawa}, \citenamefont {Itou}, \citenamefont
  {Ito}, \citenamefont {Taniguchi}, \citenamefont {Saito}, \citenamefont
  {Iguchi}, \citenamefont {Sasaki},\ and\ \citenamefont
  {Kanoda}}]{furukawa_quantum_2015}%
  \BibitemOpen
  \bibfield  {author} {\bibinfo {author} {\bibfnamefont {T.}~\bibnamefont
  {Furukawa}}, \bibinfo {author} {\bibfnamefont {K.}~\bibnamefont {Miyagawa}},
  \bibinfo {author} {\bibfnamefont {T.}~\bibnamefont {Itou}}, \bibinfo {author}
  {\bibfnamefont {M.}~\bibnamefont {Ito}}, \bibinfo {author} {\bibfnamefont
  {H.}~\bibnamefont {Taniguchi}}, \bibinfo {author} {\bibfnamefont
  {M.}~\bibnamefont {Saito}}, \bibinfo {author} {\bibfnamefont
  {S.}~\bibnamefont {Iguchi}}, \bibinfo {author} {\bibfnamefont
  {T.}~\bibnamefont {Sasaki}}, \ and\ \bibinfo {author} {\bibfnamefont
  {K.}~\bibnamefont {Kanoda}},\ }\bibfield  {title} {\enquote {\bibinfo {title}
  {Quantum {{Spin Liquid Emerging}} from {{Antiferromagnetic Order}} by
  {{Introducing Disorder}}},}\ }\href {\doibase 10.1103/PhysRevLett.115.077001}
  {\bibfield  {journal} {\bibinfo  {journal} {Phys. Rev. Lett.}\ }\textbf
  {\bibinfo {volume} {115}},\ \bibinfo {pages} {077001} (\bibinfo {year}
  {2015})}\BibitemShut {NoStop}%
\bibitem [{\citenamefont {Savary}\ and\ \citenamefont
  {Balents}(2017)}]{savary_disorder-induced_2017}%
  \BibitemOpen
  \bibfield  {author} {\bibinfo {author} {\bibfnamefont {Lucile}\ \bibnamefont
  {Savary}}\ and\ \bibinfo {author} {\bibfnamefont {Leon}\ \bibnamefont
  {Balents}},\ }\bibfield  {title} {\enquote {\bibinfo {title}
  {Disorder-{{Induced Quantum Spin Liquid}} in {{Spin Ice Pyrochlores}}},}\
  }\href {\doibase 10.1103/PhysRevLett.118.087203} {\bibfield  {journal}
  {\bibinfo  {journal} {Phys. Rev. Lett.}\ }\textbf {\bibinfo {volume} {118}},\
  \bibinfo {pages} {087203} (\bibinfo {year} {2017})}\BibitemShut {NoStop}%
\bibitem [{\citenamefont {Sickafus}\ \emph {et~al.}(1999)\citenamefont
  {Sickafus}, \citenamefont {Wills},\ and\ \citenamefont
  {Grimes}}]{sickafus_structure_1999}%
  \BibitemOpen
  \bibfield  {author} {\bibinfo {author} {\bibfnamefont {Kurt~E.}\ \bibnamefont
  {Sickafus}}, \bibinfo {author} {\bibfnamefont {John~M.}\ \bibnamefont
  {Wills}}, \ and\ \bibinfo {author} {\bibfnamefont {Norman~W.}\ \bibnamefont
  {Grimes}},\ }\bibfield  {title} {\enquote {\bibinfo {title} {Structure of
  {{Spinel}}},}\ }\href {\doibase 10.1111/j.1151-2916.1999.tb02241.x}
  {\bibfield  {journal} {\bibinfo  {journal} {J. Am. Ceram. Soc.}\ }\textbf
  {\bibinfo {volume} {82}},\ \bibinfo {pages} {3279} (\bibinfo {year}
  {1999})}\BibitemShut {NoStop}%
\bibitem [{\citenamefont {Ndione}\ \emph {et~al.}(2014)\citenamefont {Ndione},
  \citenamefont {Shi}, \citenamefont {Stevanovic}, \citenamefont {Lany},
  \citenamefont {Zakutayev}, \citenamefont {Parilla}, \citenamefont {Perkins},
  \citenamefont {Berry}, \citenamefont {Ginley},\ and\ \citenamefont
  {Toney}}]{ndione_control_2014}%
  \BibitemOpen
  \bibfield  {author} {\bibinfo {author} {\bibfnamefont {Paul~F.}\ \bibnamefont
  {Ndione}}, \bibinfo {author} {\bibfnamefont {Yezhou}\ \bibnamefont {Shi}},
  \bibinfo {author} {\bibfnamefont {Vladan}\ \bibnamefont {Stevanovic}},
  \bibinfo {author} {\bibfnamefont {Stephan}\ \bibnamefont {Lany}}, \bibinfo
  {author} {\bibfnamefont {Andriy}\ \bibnamefont {Zakutayev}}, \bibinfo
  {author} {\bibfnamefont {Philip~A.}\ \bibnamefont {Parilla}}, \bibinfo
  {author} {\bibfnamefont {John~D.}\ \bibnamefont {Perkins}}, \bibinfo {author}
  {\bibfnamefont {Joseph~J.}\ \bibnamefont {Berry}}, \bibinfo {author}
  {\bibfnamefont {David~S.}\ \bibnamefont {Ginley}}, \ and\ \bibinfo {author}
  {\bibfnamefont {Michael~F.}\ \bibnamefont {Toney}},\ }\bibfield  {title}
  {\enquote {\bibinfo {title} {Control of the {{Electrical Properties}} in
  {{Spinel Oxides}} by {{Manipulating}} the {{Cation Disorder}}},}\ }\href
  {\doibase 10.1002/adfm.201302535} {\bibfield  {journal} {\bibinfo  {journal}
  {Adv. Funct. Mater.}\ }\textbf {\bibinfo {volume} {24}},\ \bibinfo {pages}
  {610} (\bibinfo {year} {2014})}\BibitemShut {NoStop}%
\bibitem [{\citenamefont {Shen}\ \emph {et~al.}(2016)\citenamefont {Shen},
  \citenamefont {Li}, \citenamefont {Wo}, \citenamefont {Li}, \citenamefont
  {Shen}, \citenamefont {Pan}, \citenamefont {Wang}, \citenamefont {Walker},
  \citenamefont {Steffens}, \citenamefont {Boehm}, \citenamefont {Hao},
  \citenamefont {{Quintero-Castro}}, \citenamefont {Harriger}, \citenamefont
  {Frontzek}, \citenamefont {Hao}, \citenamefont {Meng}, \citenamefont {Zhang},
  \citenamefont {Chen},\ and\ \citenamefont {Zhao}}]{shen_evidence_2016}%
  \BibitemOpen
  \bibfield  {author} {\bibinfo {author} {\bibfnamefont {Yao}\ \bibnamefont
  {Shen}}, \bibinfo {author} {\bibfnamefont {Yao-Dong}\ \bibnamefont {Li}},
  \bibinfo {author} {\bibfnamefont {Hongliang}\ \bibnamefont {Wo}}, \bibinfo
  {author} {\bibfnamefont {Yuesheng}\ \bibnamefont {Li}}, \bibinfo {author}
  {\bibfnamefont {Shoudong}\ \bibnamefont {Shen}}, \bibinfo {author}
  {\bibfnamefont {Bingying}\ \bibnamefont {Pan}}, \bibinfo {author}
  {\bibfnamefont {Qisi}\ \bibnamefont {Wang}}, \bibinfo {author} {\bibfnamefont
  {H.~C.}\ \bibnamefont {Walker}}, \bibinfo {author} {\bibfnamefont
  {P.}~\bibnamefont {Steffens}}, \bibinfo {author} {\bibfnamefont
  {M.}~\bibnamefont {Boehm}}, \bibinfo {author} {\bibfnamefont {Yiqing}\
  \bibnamefont {Hao}}, \bibinfo {author} {\bibfnamefont {D.~L.}\ \bibnamefont
  {{Quintero-Castro}}}, \bibinfo {author} {\bibfnamefont {L.~W.}\ \bibnamefont
  {Harriger}}, \bibinfo {author} {\bibfnamefont {M.~D.}\ \bibnamefont
  {Frontzek}}, \bibinfo {author} {\bibfnamefont {Lijie}\ \bibnamefont {Hao}},
  \bibinfo {author} {\bibfnamefont {Siqin}\ \bibnamefont {Meng}}, \bibinfo
  {author} {\bibfnamefont {Qingming}\ \bibnamefont {Zhang}}, \bibinfo {author}
  {\bibfnamefont {Gang}\ \bibnamefont {Chen}}, \ and\ \bibinfo {author}
  {\bibfnamefont {Jun}\ \bibnamefont {Zhao}},\ }\bibfield  {title} {\enquote
  {\bibinfo {title} {Evidence for a spinon {{Fermi}} surface in a
  triangular-lattice quantum-spin-liquid candidate},}\ }\href {\doibase
  10.1038/nature20614} {\bibfield  {journal} {\bibinfo  {journal} {Nature}\
  }\textbf {\bibinfo {volume} {540}},\ \bibinfo {pages} {559} (\bibinfo {year}
  {2016})}\BibitemShut {NoStop}%
\bibitem [{\citenamefont {Gao}\ \emph {et~al.}(2019)\citenamefont {Gao},
  \citenamefont {Chen}, \citenamefont {Tam}, \citenamefont {Huang},
  \citenamefont {Sasmal}, \citenamefont {Adroja}, \citenamefont {Ye},
  \citenamefont {Cao}, \citenamefont {Sala}, \citenamefont {Stone},
  \citenamefont {Baines}, \citenamefont {Verezhak}, \citenamefont {Hu},
  \citenamefont {Chung}, \citenamefont {Xu}, \citenamefont {Cheong},
  \citenamefont {Nallaiyan}, \citenamefont {Spagna}, \citenamefont {Maple},
  \citenamefont {Nevidomskyy}, \citenamefont {Morosan}, \citenamefont {Chen},\
  and\ \citenamefont {Dai}}]{gao_experimental_2019}%
  \BibitemOpen
  \bibfield  {author} {\bibinfo {author} {\bibfnamefont {Bin}\ \bibnamefont
  {Gao}}, \bibinfo {author} {\bibfnamefont {Tong}\ \bibnamefont {Chen}},
  \bibinfo {author} {\bibfnamefont {David~W.}\ \bibnamefont {Tam}}, \bibinfo
  {author} {\bibfnamefont {Chien-Lung}\ \bibnamefont {Huang}}, \bibinfo
  {author} {\bibfnamefont {Kalyan}\ \bibnamefont {Sasmal}}, \bibinfo {author}
  {\bibfnamefont {Devashibhai~T.}\ \bibnamefont {Adroja}}, \bibinfo {author}
  {\bibfnamefont {Feng}\ \bibnamefont {Ye}}, \bibinfo {author} {\bibfnamefont
  {Huibo}\ \bibnamefont {Cao}}, \bibinfo {author} {\bibfnamefont {Gabriele}\
  \bibnamefont {Sala}}, \bibinfo {author} {\bibfnamefont {Matthew~B.}\
  \bibnamefont {Stone}}, \bibinfo {author} {\bibfnamefont {Christopher}\
  \bibnamefont {Baines}}, \bibinfo {author} {\bibfnamefont {Joel A.~T.}\
  \bibnamefont {Verezhak}}, \bibinfo {author} {\bibfnamefont {Haoyu}\
  \bibnamefont {Hu}}, \bibinfo {author} {\bibfnamefont {Jae-Ho}\ \bibnamefont
  {Chung}}, \bibinfo {author} {\bibfnamefont {Xianghan}\ \bibnamefont {Xu}},
  \bibinfo {author} {\bibfnamefont {Sang-Wook}\ \bibnamefont {Cheong}},
  \bibinfo {author} {\bibfnamefont {Manivannan}\ \bibnamefont {Nallaiyan}},
  \bibinfo {author} {\bibfnamefont {Stefano}\ \bibnamefont {Spagna}}, \bibinfo
  {author} {\bibfnamefont {M.~Brian}\ \bibnamefont {Maple}}, \bibinfo {author}
  {\bibfnamefont {Andriy~H.}\ \bibnamefont {Nevidomskyy}}, \bibinfo {author}
  {\bibfnamefont {Emilia}\ \bibnamefont {Morosan}}, \bibinfo {author}
  {\bibfnamefont {Gang}\ \bibnamefont {Chen}}, \ and\ \bibinfo {author}
  {\bibfnamefont {Pengcheng}\ \bibnamefont {Dai}},\ }\bibfield  {title}
  {\enquote {\bibinfo {title} {Experimental signatures of a three-dimensional
  quantum spin liquid in effective spin-1/2
  {{Ce}}{\textsubscript{2}}{{Zr}}{\textsubscript{2}}{{O}}{\textsubscript{7}}
  pyrochlore},}\ }\href {\doibase 10.1038/s41567-019-0577-6} {\bibfield
  {journal} {\bibinfo  {journal} {Nat. Phys.}\ }\textbf {\bibinfo {volume}
  {15}},\ \bibinfo {pages} {1052} (\bibinfo {year} {2019})}\BibitemShut
  {NoStop}%
\bibitem [{\citenamefont {Zeng}\ \emph {et~al.}(2024)\citenamefont {Zeng},
  \citenamefont {Zhou}, \citenamefont {Zhou}, \citenamefont {Han},
  \citenamefont {Chi}, \citenamefont {Li}, \citenamefont {Kofu}, \citenamefont
  {Nakajima}, \citenamefont {Wei}, \citenamefont {Zhang}, \citenamefont
  {Mazzone}, \citenamefont {Meng},\ and\ \citenamefont
  {Li}}]{zeng_spectral_2024}%
  \BibitemOpen
  \bibfield  {author} {\bibinfo {author} {\bibfnamefont {Zhenyuan}\
  \bibnamefont {Zeng}}, \bibinfo {author} {\bibfnamefont {Chengkang}\
  \bibnamefont {Zhou}}, \bibinfo {author} {\bibfnamefont {Honglin}\
  \bibnamefont {Zhou}}, \bibinfo {author} {\bibfnamefont {Lankun}\ \bibnamefont
  {Han}}, \bibinfo {author} {\bibfnamefont {Runze}\ \bibnamefont {Chi}},
  \bibinfo {author} {\bibfnamefont {Kuo}\ \bibnamefont {Li}}, \bibinfo {author}
  {\bibfnamefont {Maiko}\ \bibnamefont {Kofu}}, \bibinfo {author}
  {\bibfnamefont {Kenji}\ \bibnamefont {Nakajima}}, \bibinfo {author}
  {\bibfnamefont {Yuan}\ \bibnamefont {Wei}}, \bibinfo {author} {\bibfnamefont
  {Wenliang}\ \bibnamefont {Zhang}}, \bibinfo {author} {\bibfnamefont
  {Daniel~G.}\ \bibnamefont {Mazzone}}, \bibinfo {author} {\bibfnamefont
  {Zi~Yang}\ \bibnamefont {Meng}}, \ and\ \bibinfo {author} {\bibfnamefont
  {Shiliang}\ \bibnamefont {Li}},\ }\bibfield  {title} {\enquote {\bibinfo
  {title} {Spectral evidence for {{Dirac}} spinons in a kagome lattice
  antiferromagnet},}\ }\href {\doibase 10.1038/s41567-024-02495-z} {\bibfield
  {journal} {\bibinfo  {journal} {Nat. Phys.}\ }\textbf {\bibinfo {volume}
  {20}},\ \bibinfo {pages} {1097} (\bibinfo {year} {2024})}\BibitemShut
  {NoStop}%
\bibitem [{\citenamefont {Rodriguez}\ \emph {et~al.}(2008)\citenamefont
  {Rodriguez}, \citenamefont {Adler}, \citenamefont {Brand}, \citenamefont
  {Broholm}, \citenamefont {Cook}, \citenamefont {Brocker}, \citenamefont
  {Hammond}, \citenamefont {Huang}, \citenamefont {Hundertmark}, \citenamefont
  {Lynn}, \citenamefont {Maliszewskyj}, \citenamefont {Moyer}, \citenamefont
  {Orndorff}, \citenamefont {Pierce}, \citenamefont {Pike}, \citenamefont
  {Scharfstein}, \citenamefont {Smee},\ and\ \citenamefont
  {Vilaseca}}]{rodriguez_macsnew_2008}%
  \BibitemOpen
  \bibfield  {author} {\bibinfo {author} {\bibfnamefont {J~A}\ \bibnamefont
  {Rodriguez}}, \bibinfo {author} {\bibfnamefont {D~M}\ \bibnamefont {Adler}},
  \bibinfo {author} {\bibfnamefont {P~C}\ \bibnamefont {Brand}}, \bibinfo
  {author} {\bibfnamefont {C}~\bibnamefont {Broholm}}, \bibinfo {author}
  {\bibfnamefont {J~C}\ \bibnamefont {Cook}}, \bibinfo {author} {\bibfnamefont
  {C}~\bibnamefont {Brocker}}, \bibinfo {author} {\bibfnamefont
  {R}~\bibnamefont {Hammond}}, \bibinfo {author} {\bibfnamefont
  {Z}~\bibnamefont {Huang}}, \bibinfo {author} {\bibfnamefont {P}~\bibnamefont
  {Hundertmark}}, \bibinfo {author} {\bibfnamefont {J~W}\ \bibnamefont {Lynn}},
  \bibinfo {author} {\bibfnamefont {N~C}\ \bibnamefont {Maliszewskyj}},
  \bibinfo {author} {\bibfnamefont {J}~\bibnamefont {Moyer}}, \bibinfo {author}
  {\bibfnamefont {J}~\bibnamefont {Orndorff}}, \bibinfo {author} {\bibfnamefont
  {D}~\bibnamefont {Pierce}}, \bibinfo {author} {\bibfnamefont {T~D}\
  \bibnamefont {Pike}}, \bibinfo {author} {\bibfnamefont {G}~\bibnamefont
  {Scharfstein}}, \bibinfo {author} {\bibfnamefont {S~A}\ \bibnamefont {Smee}},
  \ and\ \bibinfo {author} {\bibfnamefont {R}~\bibnamefont {Vilaseca}},\
  }\bibfield  {title} {\enquote {\bibinfo {title} {{{MACS}}---a new high
  intensity cold neutron spectrometer at {{NIST}}},}\ }\href {\doibase
  10.1088/0957-0233/19/3/034023} {\bibfield  {journal} {\bibinfo  {journal}
  {Meas. Sci. Technol.}\ }\textbf {\bibinfo {volume} {19}},\ \bibinfo {pages}
  {34023} (\bibinfo {year} {2008})}\BibitemShut {NoStop}%
\bibitem [{\citenamefont {Granroth}\ \emph {et~al.}(2010)\citenamefont
  {Granroth}, \citenamefont {Kolesnikov}, \citenamefont {Sherline},
  \citenamefont {Clancy}, \citenamefont {Ross}, \citenamefont {Ruff},
  \citenamefont {Gaulin},\ and\ \citenamefont
  {Nagler}}]{granroth_sequoia_2010}%
  \BibitemOpen
  \bibfield  {author} {\bibinfo {author} {\bibfnamefont {G~E}\ \bibnamefont
  {Granroth}}, \bibinfo {author} {\bibfnamefont {A~I}\ \bibnamefont
  {Kolesnikov}}, \bibinfo {author} {\bibfnamefont {T~E}\ \bibnamefont
  {Sherline}}, \bibinfo {author} {\bibfnamefont {J~P}\ \bibnamefont {Clancy}},
  \bibinfo {author} {\bibfnamefont {K~A}\ \bibnamefont {Ross}}, \bibinfo
  {author} {\bibfnamefont {J~P~C}\ \bibnamefont {Ruff}}, \bibinfo {author}
  {\bibfnamefont {B~D}\ \bibnamefont {Gaulin}}, \ and\ \bibinfo {author}
  {\bibfnamefont {S~E}\ \bibnamefont {Nagler}},\ }\bibfield  {title} {\enquote
  {\bibinfo {title} {{{SEQUOIA}}: {{A Newly Operating Chopper Spectrometer}} at
  the {{SNS}}},}\ }\href {\doibase 10.1088/1742-6596/251/1/012058} {\bibfield
  {journal} {\bibinfo  {journal} {J. Phys. Conf. Ser.}\ }\textbf {\bibinfo
  {volume} {251}},\ \bibinfo {pages} {12058} (\bibinfo {year}
  {2010})}\BibitemShut {NoStop}%
\bibitem [{\citenamefont {Momma}\ and\ \citenamefont
  {Izumi}(2011)}]{momma_vesta_2011}%
  \BibitemOpen
  \bibfield  {author} {\bibinfo {author} {\bibfnamefont {Koichi}\ \bibnamefont
  {Momma}}\ and\ \bibinfo {author} {\bibfnamefont {Fujio}\ \bibnamefont
  {Izumi}},\ }\bibfield  {title} {\enquote {\bibinfo {title}
  {{{{\emph{VESTA}}}}{\emph{ 3}} for three-dimensional visualization of
  crystal, volumetric and morphology data},}\ }\href {\doibase
  10.1107/S0021889811038970} {\bibfield  {journal} {\bibinfo  {journal} {J.
  Appl. Crystallogr.}\ }\textbf {\bibinfo {volume} {44}},\ \bibinfo {pages}
  {1272} (\bibinfo {year} {2011})}\BibitemShut {NoStop}%
\bibitem [{\citenamefont {Lopitaux}\ \emph {et~al.}(1976)\citenamefont
  {Lopitaux}, \citenamefont {Arsene},\ and\ \citenamefont
  {Lenglet}}]{lopitaux_etude_1976}%
  \BibitemOpen
  \bibfield  {author} {\bibinfo {author} {\bibfnamefont {J.}~\bibnamefont
  {Lopitaux}}, \bibinfo {author} {\bibfnamefont {J.}~\bibnamefont {Arsene}}, \
  and\ \bibinfo {author} {\bibfnamefont {M.}~\bibnamefont {Lenglet}},\
  }\bibfield  {title} {\enquote {\bibinfo {title} {{Etude du syst{\`e}me
  Li\textsubscript{0,5}Ga\textsubscript{2,5}O\textsubscript{4}
  CuGa\textsubscript{2}O\textsubscript{4} par spectrom{\'e}trie IR et
  diffraction de neutrons}},}\ }\href {\doibase 10.1016/0022-1902(76)80012-7}
  {\bibfield  {journal} {\bibinfo  {journal} {J. Inorg. Nucl. Chem.}\ }\textbf
  {\bibinfo {volume} {38}},\ \bibinfo {pages} {985} (\bibinfo {year}
  {1976})}\BibitemShut {NoStop}%
\bibitem [{\citenamefont {Gonz{\'a}lez}\ and\ \citenamefont
  {Are{\'a}n}(1985)}]{gonzalez_x-ray_1985}%
  \BibitemOpen
  \bibfield  {author} {\bibinfo {author} {\bibfnamefont {Juan M.~Rubio}\
  \bibnamefont {Gonz{\'a}lez}}\ and\ \bibinfo {author} {\bibfnamefont
  {Carlos~Otero}\ \bibnamefont {Are{\'a}n}},\ }\bibfield  {title} {\enquote
  {\bibinfo {title} {X-{{Ray}} diffraction determination of the cation
  distribution and oxygen positional parameter in polycrystalline spinels},}\
  }\href {\doibase 10.1039/DT9850002155} {\bibfield  {journal} {\bibinfo
  {journal} {J. Chem. Soc. Dalton Trans.}\ }\textbf {\bibinfo {volume}
  {1985}},\ \bibinfo {pages} {2155} (\bibinfo {year} {1985})}\BibitemShut
  {NoStop}%
\bibitem [{\citenamefont {Petrakovskii}\ \emph {et~al.}(2001)\citenamefont
  {Petrakovskii}, \citenamefont {Aleksandrov}, \citenamefont {Bezmaternikh},
  \citenamefont {Aplesnin}, \citenamefont {Roessli}, \citenamefont {Semadeni},
  \citenamefont {Amato}, \citenamefont {Baines}, \citenamefont
  {Bartolom{\'e}},\ and\ \citenamefont
  {Evangelisti}}]{petrakovskii_spin-glass_2001}%
  \BibitemOpen
  \bibfield  {author} {\bibinfo {author} {\bibfnamefont {G.~A.}\ \bibnamefont
  {Petrakovskii}}, \bibinfo {author} {\bibfnamefont {K.~S.}\ \bibnamefont
  {Aleksandrov}}, \bibinfo {author} {\bibfnamefont {L.~N.}\ \bibnamefont
  {Bezmaternikh}}, \bibinfo {author} {\bibfnamefont {S.~S.}\ \bibnamefont
  {Aplesnin}}, \bibinfo {author} {\bibfnamefont {B.}~\bibnamefont {Roessli}},
  \bibinfo {author} {\bibfnamefont {F.}~\bibnamefont {Semadeni}}, \bibinfo
  {author} {\bibfnamefont {A.}~\bibnamefont {Amato}}, \bibinfo {author}
  {\bibfnamefont {C.}~\bibnamefont {Baines}}, \bibinfo {author} {\bibfnamefont
  {J.}~\bibnamefont {Bartolom{\'e}}}, \ and\ \bibinfo {author} {\bibfnamefont
  {M.}~\bibnamefont {Evangelisti}},\ }\bibfield  {title} {\enquote {\bibinfo
  {title} {Spin-glass state in
  {{CuGa}}{\textsubscript{2}}{{O}}{\textsubscript{4}}},}\ }\href {\doibase
  10.1103/PhysRevB.63.184425} {\bibfield  {journal} {\bibinfo  {journal} {Phys.
  Rev. B}\ }\textbf {\bibinfo {volume} {63}},\ \bibinfo {pages} {184425}
  (\bibinfo {year} {2001})}\BibitemShut {NoStop}%
\bibitem [{\citenamefont {Helton}\ \emph {et~al.}(2007)\citenamefont {Helton},
  \citenamefont {Matan}, \citenamefont {Shores}, \citenamefont {Nytko},
  \citenamefont {Bartlett}, \citenamefont {Yoshida}, \citenamefont {Takano},
  \citenamefont {Suslov}, \citenamefont {Qiu}, \citenamefont {Chung},
  \citenamefont {Nocera},\ and\ \citenamefont {Lee}}]{helton_spin_2007}%
  \BibitemOpen
  \bibfield  {author} {\bibinfo {author} {\bibfnamefont {J.~S.}\ \bibnamefont
  {Helton}}, \bibinfo {author} {\bibfnamefont {K.}~\bibnamefont {Matan}},
  \bibinfo {author} {\bibfnamefont {M.~P.}\ \bibnamefont {Shores}}, \bibinfo
  {author} {\bibfnamefont {E.~A.}\ \bibnamefont {Nytko}}, \bibinfo {author}
  {\bibfnamefont {B.~M.}\ \bibnamefont {Bartlett}}, \bibinfo {author}
  {\bibfnamefont {Y.}~\bibnamefont {Yoshida}}, \bibinfo {author} {\bibfnamefont
  {Y.}~\bibnamefont {Takano}}, \bibinfo {author} {\bibfnamefont
  {A.}~\bibnamefont {Suslov}}, \bibinfo {author} {\bibfnamefont
  {Y.}~\bibnamefont {Qiu}}, \bibinfo {author} {\bibfnamefont {J.-H.}\
  \bibnamefont {Chung}}, \bibinfo {author} {\bibfnamefont {D.~G.}\ \bibnamefont
  {Nocera}}, \ and\ \bibinfo {author} {\bibfnamefont {Y.~S.}\ \bibnamefont
  {Lee}},\ }\bibfield  {title} {\enquote {\bibinfo {title} {Spin {{Dynamics}}
  of the {{Spin-}}{$1/2$} {{Kagome Lattice Antiferromagnet
  ZnCu}}{\textsubscript{3}}({{OH}}){\textsubscript{6}}{{Cl}}{\textsubscript{2}}},}\
  }\href {\doibase 10.1103/PhysRevLett.98.107204} {\bibfield  {journal}
  {\bibinfo  {journal} {Phys. Rev. Lett.}\ }\textbf {\bibinfo {volume} {98}},\
  \bibinfo {pages} {107204} (\bibinfo {year} {2007})}\BibitemShut {NoStop}%
\bibitem [{\citenamefont {Yamashita}\ \emph {et~al.}(2008)\citenamefont
  {Yamashita}, \citenamefont {Nakazawa}, \citenamefont {Oguni}, \citenamefont
  {Oshima}, \citenamefont {Nojiri}, \citenamefont {Shimizu}, \citenamefont
  {Miyagawa},\ and\ \citenamefont {Kanoda}}]{yamashita_thermodynamic_2008}%
  \BibitemOpen
  \bibfield  {author} {\bibinfo {author} {\bibfnamefont {Satoshi}\ \bibnamefont
  {Yamashita}}, \bibinfo {author} {\bibfnamefont {Yasuhiro}\ \bibnamefont
  {Nakazawa}}, \bibinfo {author} {\bibfnamefont {Masaharu}\ \bibnamefont
  {Oguni}}, \bibinfo {author} {\bibfnamefont {Yugo}\ \bibnamefont {Oshima}},
  \bibinfo {author} {\bibfnamefont {Hiroyuki}\ \bibnamefont {Nojiri}}, \bibinfo
  {author} {\bibfnamefont {Yasuhiro}\ \bibnamefont {Shimizu}}, \bibinfo
  {author} {\bibfnamefont {Kazuya}\ \bibnamefont {Miyagawa}}, \ and\ \bibinfo
  {author} {\bibfnamefont {Kazushi}\ \bibnamefont {Kanoda}},\ }\bibfield
  {title} {\enquote {\bibinfo {title} {Thermodynamic properties of a spin-1/2
  spin-liquid state in a {$\kappa$}-type organic salt},}\ }\href {\doibase
  10.1038/nphys942} {\bibfield  {journal} {\bibinfo  {journal} {Nat. Phys.}\
  }\textbf {\bibinfo {volume} {4}},\ \bibinfo {pages} {459} (\bibinfo {year}
  {2008})}\BibitemShut {NoStop}%
\bibitem [{\citenamefont {Huang}\ \emph {et~al.}(2021)\citenamefont {Huang},
  \citenamefont {Xu}, \citenamefont {Wang}, \citenamefont {Zhao}, \citenamefont
  {Tu}, \citenamefont {Ni}, \citenamefont {Wang}, \citenamefont {Pan},
  \citenamefont {Fu}, \citenamefont {Hao}, \citenamefont {Liu}, \citenamefont
  {Mei},\ and\ \citenamefont {Li}}]{huang_heat_2021}%
  \BibitemOpen
  \bibfield  {author} {\bibinfo {author} {\bibfnamefont {Y.~Y.}\ \bibnamefont
  {Huang}}, \bibinfo {author} {\bibfnamefont {Y.}~\bibnamefont {Xu}}, \bibinfo
  {author} {\bibfnamefont {Le}~\bibnamefont {Wang}}, \bibinfo {author}
  {\bibfnamefont {C.~C.}\ \bibnamefont {Zhao}}, \bibinfo {author}
  {\bibfnamefont {C.~P.}\ \bibnamefont {Tu}}, \bibinfo {author} {\bibfnamefont
  {J.~M.}\ \bibnamefont {Ni}}, \bibinfo {author} {\bibfnamefont {L.~S.}\
  \bibnamefont {Wang}}, \bibinfo {author} {\bibfnamefont {B.~L.}\ \bibnamefont
  {Pan}}, \bibinfo {author} {\bibfnamefont {Ying}\ \bibnamefont {Fu}}, \bibinfo
  {author} {\bibfnamefont {Zhanyang}\ \bibnamefont {Hao}}, \bibinfo {author}
  {\bibfnamefont {Cai}\ \bibnamefont {Liu}}, \bibinfo {author} {\bibfnamefont
  {Jia-Wei}\ \bibnamefont {Mei}}, \ and\ \bibinfo {author} {\bibfnamefont
  {S.~Y.}\ \bibnamefont {Li}},\ }\bibfield  {title} {\enquote {\bibinfo {title}
  {Heat {{Transport}} in {{Herbertsmithite}}: {{Can}} a {{Quantum Spin Liquid
  Survive Disorder}}?}}\ }\href {\doibase 10.1103/PhysRevLett.127.267202}
  {\bibfield  {journal} {\bibinfo  {journal} {Phys. Rev. Lett.}\ }\textbf
  {\bibinfo {volume} {127}},\ \bibinfo {pages} {267202} (\bibinfo {year}
  {2021})}\BibitemShut {NoStop}%
\bibitem [{\citenamefont {Ma}\ \emph {et~al.}(2024)\citenamefont {Ma},
  \citenamefont {Zheng}, \citenamefont {Chen}, \citenamefont {Xu},
  \citenamefont {Dong}, \citenamefont {Wang}, \citenamefont {Du}, \citenamefont
  {Embs}, \citenamefont {Li}, \citenamefont {Li}, \citenamefont {Zhang},
  \citenamefont {Liu}, \citenamefont {Zhong}, \citenamefont {Liu},\ and\
  \citenamefont {Wen}}]{ma_possible_2024}%
  \BibitemOpen
  \bibfield  {author} {\bibinfo {author} {\bibfnamefont {Zhen}\ \bibnamefont
  {Ma}}, \bibinfo {author} {\bibfnamefont {Shuhan}\ \bibnamefont {Zheng}},
  \bibinfo {author} {\bibfnamefont {Yingqi}\ \bibnamefont {Chen}}, \bibinfo
  {author} {\bibfnamefont {Ruokai}\ \bibnamefont {Xu}}, \bibinfo {author}
  {\bibfnamefont {Zhao-Yang}\ \bibnamefont {Dong}}, \bibinfo {author}
  {\bibfnamefont {Jinghui}\ \bibnamefont {Wang}}, \bibinfo {author}
  {\bibfnamefont {Hong}\ \bibnamefont {Du}}, \bibinfo {author} {\bibfnamefont
  {Jan~Peter}\ \bibnamefont {Embs}}, \bibinfo {author} {\bibfnamefont
  {Shuaiwei}\ \bibnamefont {Li}}, \bibinfo {author} {\bibfnamefont {Yao}\
  \bibnamefont {Li}}, \bibinfo {author} {\bibfnamefont {Yongjun}\ \bibnamefont
  {Zhang}}, \bibinfo {author} {\bibfnamefont {Meifeng}\ \bibnamefont {Liu}},
  \bibinfo {author} {\bibfnamefont {Ruidan}\ \bibnamefont {Zhong}}, \bibinfo
  {author} {\bibfnamefont {Jun-Ming}\ \bibnamefont {Liu}}, \ and\ \bibinfo
  {author} {\bibfnamefont {Jinsheng}\ \bibnamefont {Wen}},\ }\bibfield  {title}
  {\enquote {\bibinfo {title} {Possible gapless quantum spin liquid behavior in
  the triangular-lattice {{Ising}} antiferromagnet
  {{PrMgAl}}{\textsubscript{11}}{{O}}{\textsubscript{19}}},}\ }\href {\doibase
  10.1103/PhysRevB.109.165143} {\bibfield  {journal} {\bibinfo  {journal}
  {Phys. Rev. B}\ }\textbf {\bibinfo {volume} {109}},\ \bibinfo {pages}
  {165143} (\bibinfo {year} {2024})}\BibitemShut {NoStop}%
\bibitem [{\citenamefont {Kimchi}\ \emph
  {et~al.}(2018{\natexlab{a}})\citenamefont {Kimchi}, \citenamefont {Nahum},\
  and\ \citenamefont {Senthil}}]{kimchi_valence_2018}%
  \BibitemOpen
  \bibfield  {author} {\bibinfo {author} {\bibfnamefont {Itamar}\ \bibnamefont
  {Kimchi}}, \bibinfo {author} {\bibfnamefont {Adam}\ \bibnamefont {Nahum}}, \
  and\ \bibinfo {author} {\bibfnamefont {T.}~\bibnamefont {Senthil}},\
  }\bibfield  {title} {\enquote {\bibinfo {title} {Valence {{Bonds}} in
  {{Random Quantum Magnets}}: {{Theory}} and {{Application}} to
  {{YbMgGaO}}{\textsubscript{4}}},}\ }\href {\doibase
  10.1103/PhysRevX.8.031028} {\bibfield  {journal} {\bibinfo  {journal} {Phys.
  Rev. X}\ }\textbf {\bibinfo {volume} {8}},\ \bibinfo {pages} {031028}
  (\bibinfo {year} {2018}{\natexlab{a}})}\BibitemShut {NoStop}%
\bibitem [{\citenamefont {Kimchi}\ \emph
  {et~al.}(2018{\natexlab{b}})\citenamefont {Kimchi}, \citenamefont
  {Sheckelton}, \citenamefont {McQueen},\ and\ \citenamefont
  {Lee}}]{kimchi_scaling_2018}%
  \BibitemOpen
  \bibfield  {author} {\bibinfo {author} {\bibfnamefont {Itamar}\ \bibnamefont
  {Kimchi}}, \bibinfo {author} {\bibfnamefont {John~P.}\ \bibnamefont
  {Sheckelton}}, \bibinfo {author} {\bibfnamefont {Tyrel~M.}\ \bibnamefont
  {McQueen}}, \ and\ \bibinfo {author} {\bibfnamefont {Patrick~A.}\
  \bibnamefont {Lee}},\ }\bibfield  {title} {\enquote {\bibinfo {title}
  {Scaling and data collapse from local moments in frustrated disordered
  quantum spin systems},}\ }\href {\doibase 10.1038/s41467-018-06800-2}
  {\bibfield  {journal} {\bibinfo  {journal} {Nat. Commun.}\ }\textbf {\bibinfo
  {volume} {9}},\ \bibinfo {pages} {4367} (\bibinfo {year}
  {2018}{\natexlab{b}})}\BibitemShut {NoStop}%
\bibitem [{\citenamefont {Peng}\ and\ \citenamefont
  {Zhang}(2024)}]{peng_dynamical_2024}%
  \BibitemOpen
  \bibfield  {author} {\bibinfo {author} {\bibfnamefont {Chen}\ \bibnamefont
  {Peng}}\ and\ \bibinfo {author} {\bibfnamefont {Long}\ \bibnamefont
  {Zhang}},\ }\bibfield  {title} {\enquote {\bibinfo {title} {Dynamical scaling
  behavior of the two-dimensional random singlet state in the random {{Q}}
  model},}\ }\href {\doibase 10.1103/PhysRevB.110.235112} {\bibfield  {journal}
  {\bibinfo  {journal} {Phys. Rev. B}\ }\textbf {\bibinfo {volume} {110}},\
  \bibinfo {pages} {235112} (\bibinfo {year} {2024})}\BibitemShut {NoStop}%
\bibitem [{\citenamefont {Han}\ \emph {et~al.}(2016)\citenamefont {Han},
  \citenamefont {Norman}, \citenamefont {Wen}, \citenamefont
  {{Rodriguez-Rivera}}, \citenamefont {Helton}, \citenamefont {Broholm},\ and\
  \citenamefont {Lee}}]{han_correlated_2016}%
  \BibitemOpen
  \bibfield  {author} {\bibinfo {author} {\bibfnamefont {Tian-Heng}\
  \bibnamefont {Han}}, \bibinfo {author} {\bibfnamefont {M.~R.}\ \bibnamefont
  {Norman}}, \bibinfo {author} {\bibfnamefont {J.-J.}\ \bibnamefont {Wen}},
  \bibinfo {author} {\bibfnamefont {Jose~A.}\ \bibnamefont
  {{Rodriguez-Rivera}}}, \bibinfo {author} {\bibfnamefont {Joel~S.}\
  \bibnamefont {Helton}}, \bibinfo {author} {\bibfnamefont {Collin}\
  \bibnamefont {Broholm}}, \ and\ \bibinfo {author} {\bibfnamefont {Young~S.}\
  \bibnamefont {Lee}},\ }\bibfield  {title} {\enquote {\bibinfo {title}
  {Correlated impurities and intrinsic spin-liquid physics in the kagome
  material herbertsmithite},}\ }\href {\doibase 10.1103/PhysRevB.94.060409}
  {\bibfield  {journal} {\bibinfo  {journal} {Phys. Rev. B}\ }\textbf {\bibinfo
  {volume} {94}},\ \bibinfo {pages} {060409} (\bibinfo {year}
  {2016})}\BibitemShut {NoStop}%
\bibitem [{\citenamefont {Ewings}\ \emph {et~al.}(2016)\citenamefont {Ewings},
  \citenamefont {Buts}, \citenamefont {Le}, \citenamefont {Van~Duijn},
  \citenamefont {Bustinduy},\ and\ \citenamefont
  {Perring}}]{ewings_horace_2016}%
  \BibitemOpen
  \bibfield  {author} {\bibinfo {author} {\bibfnamefont {R.A.}\ \bibnamefont
  {Ewings}}, \bibinfo {author} {\bibfnamefont {A.}~\bibnamefont {Buts}},
  \bibinfo {author} {\bibfnamefont {M.D.}\ \bibnamefont {Le}}, \bibinfo
  {author} {\bibfnamefont {J.}~\bibnamefont {Van~Duijn}}, \bibinfo {author}
  {\bibfnamefont {I.}~\bibnamefont {Bustinduy}}, \ and\ \bibinfo {author}
  {\bibfnamefont {T.G.}\ \bibnamefont {Perring}},\ }\bibfield  {title}
  {\enquote {\bibinfo {title} {Horace : {{Software}} for the analysis of data
  from single crystal spectroscopy experiments at time-of-flight neutron
  instruments},}\ }\href {\doibase 10.1016/j.nima.2016.07.036} {\bibfield
  {journal} {\bibinfo  {journal} {Nucl. Instrum. Methods Phys. Res. A: Accel.
  Spectrom. Detect. Assoc. Equip.}\ }\textbf {\bibinfo {volume} {834}},\
  \bibinfo {pages} {132} (\bibinfo {year} {2016})}\BibitemShut {NoStop}%
\bibitem [{\citenamefont {Binder}\ and\ \citenamefont
  {Young}(1986)}]{binder_spin_1986}%
  \BibitemOpen
  \bibfield  {author} {\bibinfo {author} {\bibfnamefont {K.}~\bibnamefont
  {Binder}}\ and\ \bibinfo {author} {\bibfnamefont {A.~P.}\ \bibnamefont
  {Young}},\ }\bibfield  {title} {\enquote {\bibinfo {title} {Spin glasses:
  {{Experimental}} facts, theoretical concepts, and open questions},}\ }\href
  {\doibase 10.1103/RevModPhys.58.801} {\bibfield  {journal} {\bibinfo
  {journal} {Rev. Mod. Phys.}\ }\textbf {\bibinfo {volume} {58}},\ \bibinfo
  {pages} {801} (\bibinfo {year} {1986})}\BibitemShut {NoStop}%
\bibitem [{\citenamefont {Fenner}\ \emph {et~al.}(2009)\citenamefont {Fenner},
  \citenamefont {Wills}, \citenamefont {Bramwell}, \citenamefont {Dahlberg},\
  and\ \citenamefont {Schiffer}}]{fenner_zero-point_2009}%
  \BibitemOpen
  \bibfield  {author} {\bibinfo {author} {\bibfnamefont {L~A}\ \bibnamefont
  {Fenner}}, \bibinfo {author} {\bibfnamefont {A~S}\ \bibnamefont {Wills}},
  \bibinfo {author} {\bibfnamefont {S~T}\ \bibnamefont {Bramwell}}, \bibinfo
  {author} {\bibfnamefont {M}~\bibnamefont {Dahlberg}}, \ and\ \bibinfo
  {author} {\bibfnamefont {P}~\bibnamefont {Schiffer}},\ }\bibfield  {title}
  {\enquote {\bibinfo {title} {Zero-point entropy of the spinel spin glasses
  {{CuGa}}{\textsubscript{2}}{{O}}{\textsubscript{4}} and
  {{CuAl}}{\textsubscript{2}}{{O}}{\textsubscript{4}}},}\ }\href {\doibase
  10.1088/1742-6596/145/1/012029} {\bibfield  {journal} {\bibinfo  {journal}
  {J. Phys. Conf. Ser.}\ }\textbf {\bibinfo {volume} {145}},\ \bibinfo {pages}
  {012029} (\bibinfo {year} {2009})}\BibitemShut {NoStop}%
\bibitem [{\citenamefont {Mydosh}(2015)}]{mydosh_spin_2015}%
  \BibitemOpen
  \bibfield  {author} {\bibinfo {author} {\bibfnamefont {J~A}\ \bibnamefont
  {Mydosh}},\ }\bibfield  {title} {\enquote {\bibinfo {title} {Spin glasses:
  redux: an updated experimental/materials survey},}\ }\href {\doibase
  10.1088/0034-4885/78/5/052501} {\bibfield  {journal} {\bibinfo  {journal}
  {Rep. Prog. Phys.}\ }\textbf {\bibinfo {volume} {78}},\ \bibinfo {pages}
  {52501} (\bibinfo {year} {2015})}\BibitemShut {NoStop}%
\bibitem [{\citenamefont {Mehlawat}\ \emph {et~al.}(2015)\citenamefont
  {Mehlawat}, \citenamefont {Sharma},\ and\ \citenamefont
  {Singh}}]{mehlawat_fragile_2015}%
  \BibitemOpen
  \bibfield  {author} {\bibinfo {author} {\bibfnamefont {Kavita}\ \bibnamefont
  {Mehlawat}}, \bibinfo {author} {\bibfnamefont {G.}~\bibnamefont {Sharma}}, \
  and\ \bibinfo {author} {\bibfnamefont {Yogesh}\ \bibnamefont {Singh}},\
  }\bibfield  {title} {\enquote {\bibinfo {title} {Fragile magnetic order in
  the honeycomb lattice {{Iridate}} {${\mathrm{Na}}_{2}{\mathrm{IrO}}_{3}$}
  revealed by magnetic impurity doping},}\ }\href {\doibase
  10.1103/PhysRevB.92.134412} {\bibfield  {journal} {\bibinfo  {journal} {Phys.
  Rev. B}\ }\textbf {\bibinfo {volume} {92}},\ \bibinfo {pages} {134412}
  (\bibinfo {year} {2015})}\BibitemShut {NoStop}%
\bibitem [{\citenamefont {Bairwa}\ \emph {et~al.}(2025)\citenamefont {Bairwa},
  \citenamefont {Bandyopadhyay}, \citenamefont {Adroja}, \citenamefont
  {Stenning}, \citenamefont {Luetkens}, \citenamefont {Hicken}, \citenamefont
  {Krieger}, \citenamefont {Cibin}, \citenamefont {Rotter}, \citenamefont
  {Rayaprol}, \citenamefont {Babu},\ and\ \citenamefont
  {Elizabeth}}]{bairwa_quantum_2025}%
  \BibitemOpen
  \bibfield  {author} {\bibinfo {author} {\bibfnamefont {Dhanpal}\ \bibnamefont
  {Bairwa}}, \bibinfo {author} {\bibfnamefont {Abhisek}\ \bibnamefont
  {Bandyopadhyay}}, \bibinfo {author} {\bibfnamefont {Devashibhai}\
  \bibnamefont {Adroja}}, \bibinfo {author} {\bibfnamefont {G.~B.~G.}\
  \bibnamefont {Stenning}}, \bibinfo {author} {\bibfnamefont {Hubertus}\
  \bibnamefont {Luetkens}}, \bibinfo {author} {\bibfnamefont {Thomas~James}\
  \bibnamefont {Hicken}}, \bibinfo {author} {\bibfnamefont {Jonas~A.}\
  \bibnamefont {Krieger}}, \bibinfo {author} {\bibfnamefont {G.}~\bibnamefont
  {Cibin}}, \bibinfo {author} {\bibfnamefont {M.}~\bibnamefont {Rotter}},
  \bibinfo {author} {\bibfnamefont {S.}~\bibnamefont {Rayaprol}}, \bibinfo
  {author} {\bibfnamefont {P.~D.}\ \bibnamefont {Babu}}, \ and\ \bibinfo
  {author} {\bibfnamefont {Suja}\ \bibnamefont {Elizabeth}},\ }\bibfield
  {title} {\enquote {\bibinfo {title} {Quantum spin liquid ground state in the
  rare-earth triangular antiferromagnet
  {${\mathrm{SmTa}}_{7}{\mathrm{O}}_{19}$}},}\ }\href {\doibase
  10.1103/PhysRevB.111.104413} {\bibfield  {journal} {\bibinfo  {journal}
  {Phys. Rev. B}\ }\textbf {\bibinfo {volume} {111}},\ \bibinfo {pages}
  {104413} (\bibinfo {year} {2025})}\BibitemShut {NoStop}%
\bibitem [{\citenamefont {Jahn}\ \emph {et~al.}(1937)\citenamefont {Jahn},
  \citenamefont {Teller},\ and\ \citenamefont {Donnan}}]{jahn_stability_1937}%
  \BibitemOpen
  \bibfield  {author} {\bibinfo {author} {\bibfnamefont {H.~A.}\ \bibnamefont
  {Jahn}}, \bibinfo {author} {\bibfnamefont {E.}~\bibnamefont {Teller}}, \ and\
  \bibinfo {author} {\bibfnamefont {Frederick~George}\ \bibnamefont {Donnan}},\
  }\bibfield  {title} {\enquote {\bibinfo {title} {Stability of polyatomic
  molecules in degenerate electronic states - {{I}}---{{Orbital}}
  degeneracy},}\ }\href {\doibase 10.1098/rspa.1937.0142} {\bibfield  {journal}
  {\bibinfo  {journal} {Proc. R. Soc. Lond., - Math. Phys. Sci.}\ }\textbf
  {\bibinfo {volume} {161}},\ \bibinfo {pages} {220} (\bibinfo {year}
  {1937})}\BibitemShut {NoStop}%
\bibitem [{\citenamefont {Murayama}\ \emph {et~al.}(2020)\citenamefont
  {Murayama}, \citenamefont {Sato}, \citenamefont {Taniguchi}, \citenamefont
  {Kurihara}, \citenamefont {Xing}, \citenamefont {Huang}, \citenamefont
  {Kasahara}, \citenamefont {Kasahara}, \citenamefont {Kimchi}, \citenamefont
  {Yoshida}, \citenamefont {Iwasa}, \citenamefont {Mizukami}, \citenamefont
  {Shibauchi}, \citenamefont {Konczykowski},\ and\ \citenamefont
  {Matsuda}}]{murayama_effect_2020}%
  \BibitemOpen
  \bibfield  {author} {\bibinfo {author} {\bibfnamefont {H.}~\bibnamefont
  {Murayama}}, \bibinfo {author} {\bibfnamefont {Y.}~\bibnamefont {Sato}},
  \bibinfo {author} {\bibfnamefont {T.}~\bibnamefont {Taniguchi}}, \bibinfo
  {author} {\bibfnamefont {R.}~\bibnamefont {Kurihara}}, \bibinfo {author}
  {\bibfnamefont {X.~Z.}\ \bibnamefont {Xing}}, \bibinfo {author}
  {\bibfnamefont {W.}~\bibnamefont {Huang}}, \bibinfo {author} {\bibfnamefont
  {S.}~\bibnamefont {Kasahara}}, \bibinfo {author} {\bibfnamefont
  {Y.}~\bibnamefont {Kasahara}}, \bibinfo {author} {\bibfnamefont
  {I.}~\bibnamefont {Kimchi}}, \bibinfo {author} {\bibfnamefont
  {M.}~\bibnamefont {Yoshida}}, \bibinfo {author} {\bibfnamefont
  {Y.}~\bibnamefont {Iwasa}}, \bibinfo {author} {\bibfnamefont
  {Y.}~\bibnamefont {Mizukami}}, \bibinfo {author} {\bibfnamefont
  {T.}~\bibnamefont {Shibauchi}}, \bibinfo {author} {\bibfnamefont
  {M.}~\bibnamefont {Konczykowski}}, \ and\ \bibinfo {author} {\bibfnamefont
  {Y.}~\bibnamefont {Matsuda}},\ }\bibfield  {title} {\enquote {\bibinfo
  {title} {Effect of quenched disorder on the quantum spin liquid state of the
  triangular-lattice antiferromagnet
  1{{{\emph{T}}}}-{{TaS}}{\textsubscript{2}}},}\ }\href {\doibase
  10.1103/PhysRevResearch.2.013099} {\bibfield  {journal} {\bibinfo  {journal}
  {Phys. Rev. Res.}\ }\textbf {\bibinfo {volume} {2}},\ \bibinfo {pages}
  {013099} (\bibinfo {year} {2020})}\BibitemShut {NoStop}%
\bibitem [{\citenamefont {Kao}\ and\ \citenamefont
  {Perkins}(2021)}]{kao_disorder_2021}%
  \BibitemOpen
  \bibfield  {author} {\bibinfo {author} {\bibfnamefont {Wen-Han}\ \bibnamefont
  {Kao}}\ and\ \bibinfo {author} {\bibfnamefont {Natalia~B.}\ \bibnamefont
  {Perkins}},\ }\bibfield  {title} {\enquote {\bibinfo {title} {Disorder upon
  disorder: {{Localization}} effects in the {{Kitaev}} spin liquid},}\ }\href
  {\doibase 10.1016/j.aop.2021.168506} {\bibfield  {journal} {\bibinfo
  {journal} {Ann. Phys.}\ }\textbf {\bibinfo {volume} {435}},\ \bibinfo {pages}
  {168506} (\bibinfo {year} {2021})}\BibitemShut {NoStop}%
\bibitem [{\citenamefont {Norman}(2016)}]{norman_colloquium_2016}%
  \BibitemOpen
  \bibfield  {author} {\bibinfo {author} {\bibfnamefont {M.~R.}\ \bibnamefont
  {Norman}},\ }\bibfield  {title} {\enquote {\bibinfo {title}
  {{\emph{Colloquium}} : {{Herbertsmithite}} and the search for the quantum
  spin liquid},}\ }\href {\doibase 10.1103/RevModPhys.88.041002} {\bibfield
  {journal} {\bibinfo  {journal} {Rev. Mod. Phys.}\ }\textbf {\bibinfo {volume}
  {88}},\ \bibinfo {pages} {041002} (\bibinfo {year} {2016})}\BibitemShut
  {NoStop}%
\bibitem [{\citenamefont {Ding}\ \emph {et~al.}(2018)\citenamefont {Ding},
  \citenamefont {Yang}, \citenamefont {Zhang}, \citenamefont {Tan},
  \citenamefont {Zhu}, \citenamefont {Chen},\ and\ \citenamefont
  {Shu}}]{ding_possible_2018}%
  \BibitemOpen
  \bibfield  {author} {\bibinfo {author} {\bibfnamefont {Zhao-Feng}\
  \bibnamefont {Ding}}, \bibinfo {author} {\bibfnamefont {Yan-Xing}\
  \bibnamefont {Yang}}, \bibinfo {author} {\bibfnamefont {Jian}\ \bibnamefont
  {Zhang}}, \bibinfo {author} {\bibfnamefont {Cheng}\ \bibnamefont {Tan}},
  \bibinfo {author} {\bibfnamefont {Zi-Hao}\ \bibnamefont {Zhu}}, \bibinfo
  {author} {\bibfnamefont {Gang}\ \bibnamefont {Chen}}, \ and\ \bibinfo
  {author} {\bibfnamefont {Lei}\ \bibnamefont {Shu}},\ }\bibfield  {title}
  {\enquote {\bibinfo {title} {Possible gapless spin liquid in the rare-earth
  kagome lattice magnet
  {{Tm}}{\textsubscript{3}}{{Sb}}{\textsubscript{3}}{{Zn}}{\textsubscript{2}}{{O}}{\textsubscript{14}}},}\
  }\href {\doibase 10.1103/PhysRevB.98.174404} {\bibfield  {journal} {\bibinfo
  {journal} {Phys. Rev. B}\ }\textbf {\bibinfo {volume} {98}},\ \bibinfo
  {pages} {174404} (\bibinfo {year} {2018})}\BibitemShut {NoStop}%
\bibitem [{\citenamefont {Ma}\ \emph {et~al.}(2020)\citenamefont {Ma},
  \citenamefont {Dong}, \citenamefont {Wu}, \citenamefont {Zhu}, \citenamefont
  {Bao}, \citenamefont {Cai}, \citenamefont {Wang}, \citenamefont {Shangguan},
  \citenamefont {Wang}, \citenamefont {Ran}, \citenamefont {Yu}, \citenamefont
  {Deng}, \citenamefont {Mole}, \citenamefont {Li}, \citenamefont {Yu},
  \citenamefont {Li},\ and\ \citenamefont {Wen}}]{ma_disorder-induced_2020}%
  \BibitemOpen
  \bibfield  {author} {\bibinfo {author} {\bibfnamefont {Zhen}\ \bibnamefont
  {Ma}}, \bibinfo {author} {\bibfnamefont {Zhao-Yang}\ \bibnamefont {Dong}},
  \bibinfo {author} {\bibfnamefont {Si}~\bibnamefont {Wu}}, \bibinfo {author}
  {\bibfnamefont {Yinghao}\ \bibnamefont {Zhu}}, \bibinfo {author}
  {\bibfnamefont {Song}\ \bibnamefont {Bao}}, \bibinfo {author} {\bibfnamefont
  {Zhengwei}\ \bibnamefont {Cai}}, \bibinfo {author} {\bibfnamefont {Wei}\
  \bibnamefont {Wang}}, \bibinfo {author} {\bibfnamefont {Yanyan}\ \bibnamefont
  {Shangguan}}, \bibinfo {author} {\bibfnamefont {Jinghui}\ \bibnamefont
  {Wang}}, \bibinfo {author} {\bibfnamefont {Kejing}\ \bibnamefont {Ran}},
  \bibinfo {author} {\bibfnamefont {Dehong}\ \bibnamefont {Yu}}, \bibinfo
  {author} {\bibfnamefont {Guochu}\ \bibnamefont {Deng}}, \bibinfo {author}
  {\bibfnamefont {Richard~A.}\ \bibnamefont {Mole}}, \bibinfo {author}
  {\bibfnamefont {Hai-Feng}\ \bibnamefont {Li}}, \bibinfo {author}
  {\bibfnamefont {Shun-Li}\ \bibnamefont {Yu}}, \bibinfo {author}
  {\bibfnamefont {Jian-Xin}\ \bibnamefont {Li}}, \ and\ \bibinfo {author}
  {\bibfnamefont {Jinsheng}\ \bibnamefont {Wen}},\ }\bibfield  {title}
  {\enquote {\bibinfo {title} {Disorder-induced spin-liquid-like behavior in
  kagome-lattice compounds},}\ }\href {\doibase 10.1103/PhysRevB.102.224415}
  {\bibfield  {journal} {\bibinfo  {journal} {Phys. Rev. B}\ }\textbf {\bibinfo
  {volume} {102}},\ \bibinfo {pages} {224415} (\bibinfo {year}
  {2020})}\BibitemShut {NoStop}%
\bibitem [{\citenamefont {Gao}\ \emph {et~al.}(2023)\citenamefont {Gao},
  \citenamefont {Chen}, \citenamefont {Huang}, \citenamefont {Qiu},
  \citenamefont {Xu}, \citenamefont {Liebman}, \citenamefont {Chen},
  \citenamefont {Stone}, \citenamefont {Feng}, \citenamefont {Cao},
  \citenamefont {Wang}, \citenamefont {Xu}, \citenamefont {Cheong},
  \citenamefont {Winter},\ and\ \citenamefont
  {Dai}}]{gao_disorder-induced_2023}%
  \BibitemOpen
  \bibfield  {author} {\bibinfo {author} {\bibfnamefont {Bin}\ \bibnamefont
  {Gao}}, \bibinfo {author} {\bibfnamefont {Tong}\ \bibnamefont {Chen}},
  \bibinfo {author} {\bibfnamefont {Chien-Lung}\ \bibnamefont {Huang}},
  \bibinfo {author} {\bibfnamefont {Yiming}\ \bibnamefont {Qiu}}, \bibinfo
  {author} {\bibfnamefont {Guangyong}\ \bibnamefont {Xu}}, \bibinfo {author}
  {\bibfnamefont {Jesse}\ \bibnamefont {Liebman}}, \bibinfo {author}
  {\bibfnamefont {Lebing}\ \bibnamefont {Chen}}, \bibinfo {author}
  {\bibfnamefont {Matthew~B.}\ \bibnamefont {Stone}}, \bibinfo {author}
  {\bibfnamefont {Erxi}\ \bibnamefont {Feng}}, \bibinfo {author} {\bibfnamefont
  {Huibo}\ \bibnamefont {Cao}}, \bibinfo {author} {\bibfnamefont {Xiaoping}\
  \bibnamefont {Wang}}, \bibinfo {author} {\bibfnamefont {Xianghan}\
  \bibnamefont {Xu}}, \bibinfo {author} {\bibfnamefont {Sang-Wook}\
  \bibnamefont {Cheong}}, \bibinfo {author} {\bibfnamefont {Stephen~M.}\
  \bibnamefont {Winter}}, \ and\ \bibinfo {author} {\bibfnamefont {Pengcheng}\
  \bibnamefont {Dai}},\ }\bibfield  {title} {\enquote {\bibinfo {title}
  {Disorder-induced excitation continuum in a spin-1/2 cobaltate on a
  triangular lattice},}\ }\href {\doibase 10.1103/PhysRevB.108.024431}
  {\bibfield  {journal} {\bibinfo  {journal} {Phys. Rev. B}\ }\textbf {\bibinfo
  {volume} {108}},\ \bibinfo {pages} {024431} (\bibinfo {year}
  {2023})}\BibitemShut {NoStop}%
\bibitem [{\citenamefont {Shen}\ \emph {et~al.}(2018)\citenamefont {Shen},
  \citenamefont {Li}, \citenamefont {Walker}, \citenamefont {Steffens},
  \citenamefont {Boehm}, \citenamefont {Zhang}, \citenamefont {Shen},
  \citenamefont {Wo}, \citenamefont {Chen},\ and\ \citenamefont
  {Zhao}}]{shen_fractionalized_2018}%
  \BibitemOpen
  \bibfield  {author} {\bibinfo {author} {\bibfnamefont {Yao}\ \bibnamefont
  {Shen}}, \bibinfo {author} {\bibfnamefont {Yao-Dong}\ \bibnamefont {Li}},
  \bibinfo {author} {\bibfnamefont {H.~C.}\ \bibnamefont {Walker}}, \bibinfo
  {author} {\bibfnamefont {P.}~\bibnamefont {Steffens}}, \bibinfo {author}
  {\bibfnamefont {M.}~\bibnamefont {Boehm}}, \bibinfo {author} {\bibfnamefont
  {Xiaowen}\ \bibnamefont {Zhang}}, \bibinfo {author} {\bibfnamefont
  {Shoudong}\ \bibnamefont {Shen}}, \bibinfo {author} {\bibfnamefont
  {Hongliang}\ \bibnamefont {Wo}}, \bibinfo {author} {\bibfnamefont {Gang}\
  \bibnamefont {Chen}}, \ and\ \bibinfo {author} {\bibfnamefont {Jun}\
  \bibnamefont {Zhao}},\ }\bibfield  {title} {\enquote {\bibinfo {title}
  {Fractionalized excitations in the partially magnetized spin liquid candidate
  {{YbMgGaO}}{\textsubscript{4}}},}\ }\href {\doibase
  10.1038/s41467-018-06588-1} {\bibfield  {journal} {\bibinfo  {journal} {Nat.
  Commun.}\ }\textbf {\bibinfo {volume} {9}},\ \bibinfo {pages} {4138}
  (\bibinfo {year} {2018})}\BibitemShut {NoStop}%
\bibitem [{\citenamefont {Ma}\ \emph {et~al.}(2018)\citenamefont {Ma},
  \citenamefont {Wang}, \citenamefont {Dong}, \citenamefont {Zhang},
  \citenamefont {Li}, \citenamefont {Zheng}, \citenamefont {Yu}, \citenamefont
  {Wang}, \citenamefont {Che}, \citenamefont {Ran}, \citenamefont {Bao},
  \citenamefont {Cai}, \citenamefont {{\v C}erm{\'a}k}, \citenamefont
  {Schneidewind}, \citenamefont {Yano}, \citenamefont {Gardner}, \citenamefont
  {Lu}, \citenamefont {Yu}, \citenamefont {Liu}, \citenamefont {Li},
  \citenamefont {Li},\ and\ \citenamefont {Wen}}]{ma_spin-glass_2018}%
  \BibitemOpen
  \bibfield  {author} {\bibinfo {author} {\bibfnamefont {Zhen}\ \bibnamefont
  {Ma}}, \bibinfo {author} {\bibfnamefont {Jinghui}\ \bibnamefont {Wang}},
  \bibinfo {author} {\bibfnamefont {Zhao-Yang}\ \bibnamefont {Dong}}, \bibinfo
  {author} {\bibfnamefont {Jun}\ \bibnamefont {Zhang}}, \bibinfo {author}
  {\bibfnamefont {Shichao}\ \bibnamefont {Li}}, \bibinfo {author}
  {\bibfnamefont {Shu-Han}\ \bibnamefont {Zheng}}, \bibinfo {author}
  {\bibfnamefont {Yunjie}\ \bibnamefont {Yu}}, \bibinfo {author} {\bibfnamefont
  {Wei}\ \bibnamefont {Wang}}, \bibinfo {author} {\bibfnamefont {Liqiang}\
  \bibnamefont {Che}}, \bibinfo {author} {\bibfnamefont {Kejing}\ \bibnamefont
  {Ran}}, \bibinfo {author} {\bibfnamefont {Song}\ \bibnamefont {Bao}},
  \bibinfo {author} {\bibfnamefont {Zhengwei}\ \bibnamefont {Cai}}, \bibinfo
  {author} {\bibfnamefont {P.}~\bibnamefont {{\v C}erm{\'a}k}}, \bibinfo
  {author} {\bibfnamefont {A.}~\bibnamefont {Schneidewind}}, \bibinfo {author}
  {\bibfnamefont {S.}~\bibnamefont {Yano}}, \bibinfo {author} {\bibfnamefont
  {J.~S.}\ \bibnamefont {Gardner}}, \bibinfo {author} {\bibfnamefont {Xin}\
  \bibnamefont {Lu}}, \bibinfo {author} {\bibfnamefont {Shun-Li}\ \bibnamefont
  {Yu}}, \bibinfo {author} {\bibfnamefont {Jun-Ming}\ \bibnamefont {Liu}},
  \bibinfo {author} {\bibfnamefont {Shiyan}\ \bibnamefont {Li}}, \bibinfo
  {author} {\bibfnamefont {Jian-Xin}\ \bibnamefont {Li}}, \ and\ \bibinfo
  {author} {\bibfnamefont {Jinsheng}\ \bibnamefont {Wen}},\ }\bibfield  {title}
  {\enquote {\bibinfo {title} {Spin-{{Glass Ground State}} in a
  {{Triangular-Lattice Compound YbZnGaO}}{\textsubscript{4}}},}\ }\href
  {\doibase 10.1103/PhysRevLett.120.087201} {\bibfield  {journal} {\bibinfo
  {journal} {Phys. Rev. Lett.}\ }\textbf {\bibinfo {volume} {120}},\ \bibinfo
  {pages} {087201} (\bibinfo {year} {2018})}\BibitemShut {NoStop}%
\bibitem [{\citenamefont {Ma}\ \emph {et~al.}(2021)\citenamefont {Ma},
  \citenamefont {Dong}, \citenamefont {Wang}, \citenamefont {Zheng},
  \citenamefont {Ran}, \citenamefont {Bao}, \citenamefont {Cai}, \citenamefont
  {Shangguan}, \citenamefont {Wang}, \citenamefont {Boehm}, \citenamefont
  {Steffens}, \citenamefont {Regnault}, \citenamefont {Wang}, \citenamefont
  {Su}, \citenamefont {Yu}, \citenamefont {Liu}, \citenamefont {Li},\ and\
  \citenamefont {Wen}}]{ma_disorder-induced_2021}%
  \BibitemOpen
  \bibfield  {author} {\bibinfo {author} {\bibfnamefont {Zhen}\ \bibnamefont
  {Ma}}, \bibinfo {author} {\bibfnamefont {Zhao-Yang}\ \bibnamefont {Dong}},
  \bibinfo {author} {\bibfnamefont {Jinghui}\ \bibnamefont {Wang}}, \bibinfo
  {author} {\bibfnamefont {Shuhan}\ \bibnamefont {Zheng}}, \bibinfo {author}
  {\bibfnamefont {Kejing}\ \bibnamefont {Ran}}, \bibinfo {author}
  {\bibfnamefont {Song}\ \bibnamefont {Bao}}, \bibinfo {author} {\bibfnamefont
  {Zhengwei}\ \bibnamefont {Cai}}, \bibinfo {author} {\bibfnamefont {Yanyan}\
  \bibnamefont {Shangguan}}, \bibinfo {author} {\bibfnamefont {Wei}\
  \bibnamefont {Wang}}, \bibinfo {author} {\bibfnamefont {M.}~\bibnamefont
  {Boehm}}, \bibinfo {author} {\bibfnamefont {P.}~\bibnamefont {Steffens}},
  \bibinfo {author} {\bibfnamefont {L.-P.}\ \bibnamefont {Regnault}}, \bibinfo
  {author} {\bibfnamefont {Xiao}\ \bibnamefont {Wang}}, \bibinfo {author}
  {\bibfnamefont {Yixi}\ \bibnamefont {Su}}, \bibinfo {author} {\bibfnamefont
  {Shun-Li}\ \bibnamefont {Yu}}, \bibinfo {author} {\bibfnamefont {Jun-Ming}\
  \bibnamefont {Liu}}, \bibinfo {author} {\bibfnamefont {Jian-Xin}\
  \bibnamefont {Li}}, \ and\ \bibinfo {author} {\bibfnamefont {Jinsheng}\
  \bibnamefont {Wen}},\ }\bibfield  {title} {\enquote {\bibinfo {title}
  {Disorder-induced broadening of the spin waves in the triangular-lattice
  quantum spin liquid candidate {{YbZnGaO}}{\textsubscript{4}}},}\ }\href
  {\doibase 10.1103/PhysRevB.104.224433} {\bibfield  {journal} {\bibinfo
  {journal} {Phys. Rev. B}\ }\textbf {\bibinfo {volume} {104}},\ \bibinfo
  {pages} {224433} (\bibinfo {year} {2021})}\BibitemShut {NoStop}%
\bibitem [{\citenamefont {Zhu}\ \emph {et~al.}(2017)\citenamefont {Zhu},
  \citenamefont {Maksimov}, \citenamefont {White},\ and\ \citenamefont
  {Chernyshev}}]{zhu_disorder-induced_2017}%
  \BibitemOpen
  \bibfield  {author} {\bibinfo {author} {\bibfnamefont {Zhenyue}\ \bibnamefont
  {Zhu}}, \bibinfo {author} {\bibfnamefont {P.~A.}\ \bibnamefont {Maksimov}},
  \bibinfo {author} {\bibfnamefont {Steven~R.}\ \bibnamefont {White}}, \ and\
  \bibinfo {author} {\bibfnamefont {A.~L.}\ \bibnamefont {Chernyshev}},\
  }\bibfield  {title} {\enquote {\bibinfo {title} {Disorder-{{Induced Mimicry}}
  of a {{Spin Liquid}} in {{YbMgGaO}}{\textsubscript{4}}},}\ }\href {\doibase
  10.1103/PhysRevLett.119.157201} {\bibfield  {journal} {\bibinfo  {journal}
  {Phys. Rev. Lett.}\ }\textbf {\bibinfo {volume} {119}},\ \bibinfo {pages}
  {157201} (\bibinfo {year} {2017})}\BibitemShut {NoStop}%
\bibitem [{\citenamefont {Bullard}\ \emph {et~al.}(2021)\citenamefont
  {Bullard}, \citenamefont {Susner}, \citenamefont {Taddei}, \citenamefont
  {Brant},\ and\ \citenamefont {Haugan}}]{bullard_magnetic_2021}%
  \BibitemOpen
  \bibfield  {author} {\bibinfo {author} {\bibfnamefont {T.~J.}\ \bibnamefont
  {Bullard}}, \bibinfo {author} {\bibfnamefont {M.~A.}\ \bibnamefont {Susner}},
  \bibinfo {author} {\bibfnamefont {K.~M.}\ \bibnamefont {Taddei}}, \bibinfo
  {author} {\bibfnamefont {J.~A.}\ \bibnamefont {Brant}}, \ and\ \bibinfo
  {author} {\bibfnamefont {T.~J.}\ \bibnamefont {Haugan}},\ }\bibfield  {title}
  {\enquote {\bibinfo {title} {Magnetic and structural properties of the solid
  solution
  {{CuAl}}{\textsubscript{2(1-x)}}{{Ga}}{\textsubscript{2x}}{{O}}{\textsubscript{4}}},}\
  }\href {\doibase 10.1038/s41598-021-89197-1} {\bibfield  {journal} {\bibinfo
  {journal} {Sci. Rep.}\ }\textbf {\bibinfo {volume} {11}},\ \bibinfo {pages}
  {11355} (\bibinfo {year} {2021})}\BibitemShut {NoStop}%
\bibitem [{\citenamefont {Shannon}(1976)}]{shannon_revised_1976}%
  \BibitemOpen
  \bibfield  {author} {\bibinfo {author} {\bibfnamefont {R.~D.}\ \bibnamefont
  {Shannon}},\ }\bibfield  {title} {\enquote {\bibinfo {title} {Revised
  effective ionic radii and systematic studies of interatomic distances in
  halides and chalcogenides},}\ }\href {\doibase 10.1107/S0567739476001551}
  {\bibfield  {journal} {\bibinfo  {journal} {Acta Crystallogr. A}\ }\textbf
  {\bibinfo {volume} {32}},\ \bibinfo {pages} {751} (\bibinfo {year}
  {1976})}\BibitemShut {NoStop}%
\bibitem [{\citenamefont {Liu}\ \emph {et~al.}(2018)\citenamefont {Liu},
  \citenamefont {Zhang}, \citenamefont {Ji}, \citenamefont {Liu}, \citenamefont
  {Li}, \citenamefont {Wang}, \citenamefont {Lei}, \citenamefont {Chen},\ and\
  \citenamefont {Zhang}}]{liu_rare-earth_2018}%
  \BibitemOpen
  \bibfield  {author} {\bibinfo {author} {\bibfnamefont {Weiwei}\ \bibnamefont
  {Liu}}, \bibinfo {author} {\bibfnamefont {Zheng}\ \bibnamefont {Zhang}},
  \bibinfo {author} {\bibfnamefont {Jianting}\ \bibnamefont {Ji}}, \bibinfo
  {author} {\bibfnamefont {Yixuan}\ \bibnamefont {Liu}}, \bibinfo {author}
  {\bibfnamefont {Jianshu}\ \bibnamefont {Li}}, \bibinfo {author}
  {\bibfnamefont {Xiaoqun}\ \bibnamefont {Wang}}, \bibinfo {author}
  {\bibfnamefont {Hechang}\ \bibnamefont {Lei}}, \bibinfo {author}
  {\bibfnamefont {Gang}\ \bibnamefont {Chen}}, \ and\ \bibinfo {author}
  {\bibfnamefont {Qingming}\ \bibnamefont {Zhang}},\ }\bibfield  {title}
  {\enquote {\bibinfo {title} {Rare-{{Earth Chalcogenides}}: {{A Large Family}}
  of {{Triangular Lattice Spin Liquid Candidates}}},}\ }\href {\doibase
  10.1088/0256-307X/35/11/117501} {\bibfield  {journal} {\bibinfo  {journal}
  {Chin. Phys. Lett.}\ }\textbf {\bibinfo {volume} {35}},\ \bibinfo {pages}
  {117501} (\bibinfo {year} {2018})}\BibitemShut {NoStop}%
\bibitem [{\citenamefont {Bordelon}\ \emph {et~al.}(2019)\citenamefont
  {Bordelon}, \citenamefont {Kenney}, \citenamefont {Liu}, \citenamefont
  {Hogan}, \citenamefont {Posthuma}, \citenamefont {Kavand}, \citenamefont
  {Lyu}, \citenamefont {Sherwin}, \citenamefont {Butch}, \citenamefont {Brown},
  \citenamefont {Graf}, \citenamefont {Balents},\ and\ \citenamefont
  {Wilson}}]{bordelon_field-tunable_2019}%
  \BibitemOpen
  \bibfield  {author} {\bibinfo {author} {\bibfnamefont {Mitchell~M.}\
  \bibnamefont {Bordelon}}, \bibinfo {author} {\bibfnamefont {Eric}\
  \bibnamefont {Kenney}}, \bibinfo {author} {\bibfnamefont {Chunxiao}\
  \bibnamefont {Liu}}, \bibinfo {author} {\bibfnamefont {Tom}\ \bibnamefont
  {Hogan}}, \bibinfo {author} {\bibfnamefont {Lorenzo}\ \bibnamefont
  {Posthuma}}, \bibinfo {author} {\bibfnamefont {Marzieh}\ \bibnamefont
  {Kavand}}, \bibinfo {author} {\bibfnamefont {Yuanqi}\ \bibnamefont {Lyu}},
  \bibinfo {author} {\bibfnamefont {Mark}\ \bibnamefont {Sherwin}}, \bibinfo
  {author} {\bibfnamefont {N.~P.}\ \bibnamefont {Butch}}, \bibinfo {author}
  {\bibfnamefont {Craig}\ \bibnamefont {Brown}}, \bibinfo {author}
  {\bibfnamefont {M.~J.}\ \bibnamefont {Graf}}, \bibinfo {author}
  {\bibfnamefont {Leon}\ \bibnamefont {Balents}}, \ and\ \bibinfo {author}
  {\bibfnamefont {Stephen~D.}\ \bibnamefont {Wilson}},\ }\bibfield  {title}
  {\enquote {\bibinfo {title} {Field-tunable quantum disordered ground state in
  the triangular-lattice antiferromagnet {{NaYbO}}{\textsubscript{2}}},}\
  }\href {\doibase 10.1038/s41567-019-0594-5} {\bibfield  {journal} {\bibinfo
  {journal} {Nat. Phys.}\ }\textbf {\bibinfo {volume} {15}},\ \bibinfo {pages}
  {1058} (\bibinfo {year} {2019})}\BibitemShut {NoStop}%
\bibitem [{\citenamefont {Dai}\ \emph {et~al.}(2021)\citenamefont {Dai},
  \citenamefont {Zhang}, \citenamefont {Xie}, \citenamefont {Duan},
  \citenamefont {Gao}, \citenamefont {Zhu}, \citenamefont {Feng}, \citenamefont
  {Tao}, \citenamefont {Huang}, \citenamefont {Cao}, \citenamefont
  {Podlesnyak}, \citenamefont {Granroth}, \citenamefont {Everett},
  \citenamefont {Neuefeind}, \citenamefont {Voneshen}, \citenamefont {Wang},
  \citenamefont {Tan}, \citenamefont {Morosan}, \citenamefont {Wang},
  \citenamefont {Lin}, \citenamefont {Shu}, \citenamefont {Chen}, \citenamefont
  {Guo}, \citenamefont {Lu},\ and\ \citenamefont {Dai}}]{dai_spinon_2021}%
  \BibitemOpen
  \bibfield  {author} {\bibinfo {author} {\bibfnamefont {Peng-Ling}\
  \bibnamefont {Dai}}, \bibinfo {author} {\bibfnamefont {Gaoning}\ \bibnamefont
  {Zhang}}, \bibinfo {author} {\bibfnamefont {Yaofeng}\ \bibnamefont {Xie}},
  \bibinfo {author} {\bibfnamefont {Chunruo}\ \bibnamefont {Duan}}, \bibinfo
  {author} {\bibfnamefont {Yonghao}\ \bibnamefont {Gao}}, \bibinfo {author}
  {\bibfnamefont {Zihao}\ \bibnamefont {Zhu}}, \bibinfo {author} {\bibfnamefont
  {Erxi}\ \bibnamefont {Feng}}, \bibinfo {author} {\bibfnamefont {Zhen}\
  \bibnamefont {Tao}}, \bibinfo {author} {\bibfnamefont {Chien-Lung}\
  \bibnamefont {Huang}}, \bibinfo {author} {\bibfnamefont {Huibo}\ \bibnamefont
  {Cao}}, \bibinfo {author} {\bibfnamefont {Andrey}\ \bibnamefont
  {Podlesnyak}}, \bibinfo {author} {\bibfnamefont {Garrett~E.}\ \bibnamefont
  {Granroth}}, \bibinfo {author} {\bibfnamefont {Michelle~S.}\ \bibnamefont
  {Everett}}, \bibinfo {author} {\bibfnamefont {Joerg~C.}\ \bibnamefont
  {Neuefeind}}, \bibinfo {author} {\bibfnamefont {David}\ \bibnamefont
  {Voneshen}}, \bibinfo {author} {\bibfnamefont {Shun}\ \bibnamefont {Wang}},
  \bibinfo {author} {\bibfnamefont {Guotai}\ \bibnamefont {Tan}}, \bibinfo
  {author} {\bibfnamefont {Emilia}\ \bibnamefont {Morosan}}, \bibinfo {author}
  {\bibfnamefont {Xia}\ \bibnamefont {Wang}}, \bibinfo {author} {\bibfnamefont
  {Hai-Qing}\ \bibnamefont {Lin}}, \bibinfo {author} {\bibfnamefont {Lei}\
  \bibnamefont {Shu}}, \bibinfo {author} {\bibfnamefont {Gang}\ \bibnamefont
  {Chen}}, \bibinfo {author} {\bibfnamefont {Yanfeng}\ \bibnamefont {Guo}},
  \bibinfo {author} {\bibfnamefont {Xingye}\ \bibnamefont {Lu}}, \ and\
  \bibinfo {author} {\bibfnamefont {Pengcheng}\ \bibnamefont {Dai}},\
  }\bibfield  {title} {\enquote {\bibinfo {title} {Spinon {{Fermi Surface Spin
  Liquid}} in a {{Triangular Lattice Antiferromagnet}}
  {${\mathrm{NaYbSe}}_{2}$}},}\ }\href {\doibase 10.1103/PhysRevX.11.021044}
  {\bibfield  {journal} {\bibinfo  {journal} {Phys. Rev. X}\ }\textbf {\bibinfo
  {volume} {11}},\ \bibinfo {pages} {21044} (\bibinfo {year}
  {2021})}\BibitemShut {NoStop}%
\end{thebibliography}
%

\end{document}